\documentclass{LMCS}
\usepackage{times}
\usepackage{gdefs}
\usepackage{graph}
\usepackage{alltt}
\usepackage{ifthen}
\usepackage{times}
\usepackage{paralist}
\usepackage{graphics}
\usepackage{graphicx}
\usepackage{latexsym}
\usepackage{amssymb}
\usepackage{capt-of}

\usepackage{enumerate,hyperref}

\newcommand{\qE}{\quad \equiv\quad}
\newcommand{\slra}{\;\leftrightarrow\;}
\newcommand{\proves}{\vdash}
\newcommand\N{\mbox{\bf N}}    

\newcommand{\qra}{\; \rightarrow\;}
\newcommand{\half}{\frac{1}{2}}

\newcommand{\figref}[1]{Fig.~\ref{Fi:#1}}
\newcommand{\tableref}[1]{Table~\ref{Ta:#1}}

\newcommand{\secref}[1]{Section~\ref{Se:#1}}
\newcommand{\equref}[1]{Eq.~(\ref{eq:#1})}

\newcommand{\propref}[1]{Proposition~\ref{Prop:#1}}

\newcommand{\appref}[1]{Section~\ref{Se:#1}}

\newcommand{\gammaHat}{\widehat{\gamma}}

\newcommand\cA{\mathcal{A}}        

\newenvironment{theorem}{\begin{thm}}{\end{thm}}
\newenvironment{definition}{\begin{defi}}{\end{defi}}
\newenvironment{corollary}{\begin{cor}}{\end{cor}}
\newenvironment{lemma}{\begin{lem}}{\end{lem}}
\newenvironment{proposition}{\begin{prop}}{\end{prop}}
{\par\noindent{\bf Corollary~\ref{#1}\/}\begin{em}}%
{\end{em}}

\newcommand{\fo}{\mbox{{\rm FO}}}

\newcommand{\sE}{\; \equiv\;}
\newcommand{\sland}{\;\land\;}
\newcommand{\usetc}[1]{{\rm TC}[#1]}
\newcommand{\tc}{{\rm TC}}
\newcommand{\stc}[1]{#1_{\rm tc}}
\newcommand{\fotc}{{\rm FO(TC)}}
\newcommand{\abs}[1]{ \vert #1 \vert }

\newcommand{\set}[1]{\{  #1 \} }

\newcommand{\Bc}{\mathcal{B}}        
\newcommand{\cM}{\mathcal{M}}    
\newcommand{\Nc}{\mathcal{N}}    

\newcommand{\qLra}{\quad\Leftrightarrow\quad}

\newcommand{\RE}{\mbox{{\rm r.e.}}}

\renewcommand{\phi}{\varphi}       
\renewcommand{\angle}[1]{ \langle #1 \rangle }

\newcommand{\subsubsubsection}[1]{\emph{#1}:}

\newcommand{\treverse}{reverse}
\newcommand{\tappend}{append}

\newcommand{\ind}{\mbox{{\bf IND}}}  
\newcommand{\noexit}{\mbox{{\bf NoExit}}}  
\newcommand{\OUT}{\mbox{{\bf NoExit}}}  
\newcommand{\oUT}{\mbox{{\small\bf NoExit}}}  
\newcommand{\goout}{\mbox{{\bf GoOut}}}  
\newcommand{\SEP}{\mbox{{\bf GoOut}}}  
\newcommand{\sEP}{\mbox{{\small\bf GoOut}}}  
\newcommand{\newstart}{\mbox{{\bf NewStart}}}  
\newcommand{\NC}{\mbox{{\bf NewStart}}}  
\newcommand{\nC}{\mbox{{\small\bf NewStart}}}  
\newcommand{\AXIOM}[1]{{#1}}   
\newcommand{\PREMISE}[1]{P_{#1}}
\newcommand{\CONJ}[1]{C_{#1}}
\newcommand{\toprove}{\chi}
\newcommand{\premises}{\psi}
\newcommand{\conclusion}{\varphi}
\newcommand{\axioms}{\Sigma}
\newcommand{\trans}{\mbox{{\bf Trans}}} 
\newcommand{\order}{\mbox{{\bf Order}}}  
\newcommand{\NATURAL}{T_1}
\newcommand{\func}{\mbox{{\bf Func}}}  
\newcommand{\sra}{\;\rightarrow\;}
\newcommand{\tSpass}{\textsc{Spass}}
\newcommand{\Tleft}{T'_1}

\newcommand{\xbefore}{x}
\newcommand{\ybefore}{y}
\newcommand{\nbefore}{n}
\newcommand{\xentry}{xe}

\newcommand{\nentry}{ne}
\newcommand{\xafter}{x'}
\newcommand{\yafter}{y'}
\newcommand{\nafter}{n'}

\newcommand{\FUReach}[2]{r_{#1,#2}}
\newcommand{\BUReach}[2]{r_{#1,\overleftarrow{#2}}}

\newcommand{\tuple}[1]{\langle #1 \rangle}
\newcommand{\XS}{\mbox{XS}}

\newcommand{\Omit}[1]{}
\newcommand{\OnlyPaper}[1]{}

\input{xy}
\xyoption{all} 

\def\doi{5 (2:12) 2009}
\lmcsheading%
{\doi}
{1--30}
{}
{}
{Apr.~\phantom{0}2, 2006}
{May\phantom{.}~28, 2009}
{}   

\begin{document}
\title[Simulating Reachability using First-Order Logic]{Simulating Reachability using First-Order Logic
       with \\ Applications to Verification
       of Linked Data Structures\rsuper*}
\titlecomment{{\lsuper*}A preliminary version of this paper appeared in Automated Deduction -
CADE-20, 20th International Conference on Automated Deduction, Tallinn,
Estonia, July 22-27, 2005.}
\author[T.~Lev-Ami]{Tal Lev-Ami\rsuper a}%
\address{{\lsuper{a,d,f}}School of Computer Science,  Tel Aviv University}
\email{tal.levami@cs.tau.ac.il, \{msagiv,gretay\}@post.tau.ac.il}
\thanks{{\lsuper a}This research was supported by
an Adams Fellowship through the Israel Academy of Sciences and Humanities.}
\author[N.~Immerman]{Neil Immerman\rsuper b}
\address{{\lsuper{b,e}}Department of Computer Science, University of Massachusetts, Amherst}
\email{\{immerman,siddharth\}@cs.umass.edu}
\thanks{{\lsuper{b,e}}Supported by NSF grants  CCF-{0514621,0541018,0830174}.}
\author[T.~Reps]{Thomas W.~Reps\rsuper c}
\address{{\lsuper c}Computer Science Department, University of Wisconsin, Madison}
\email{reps@cs.wisc.edu}
\thanks{{\lsuper c}Supported by ONR under contracts N00014-01-1-\{0796,0708\}.}
\author[M.~Sagiv]{Mooly Sagiv\rsuper d}
\address{\vskip-6 pt}
\author[S.~Srivastava]{Siddharth Srivastava\rsuper e}
\address{\vskip-6 pt}
\author[G.~Yorsh]{Greta Yorsh\rsuper f}
\address{\vskip-6 pt}
\thanks{{\lsuper f}Partially supported by the Israeli Academy of Science}


\begin{abstract}
This paper shows how to harness existing theorem provers for
first-order logic to automatically verify safety properties of
imperative programs that perform dynamic storage allocation and
destructive updating of pointer-valued structure fields.  One of the
main obstacles is specifying and proving the (absence) of reachability
properties among dynamically allocated cells.

The main technical contributions are methods for simulating reachability in a
conservative way using first-order formulas---%
the formulas describe a superset of the set of program states that would be specified if one had a precise way to
express reachability.
\Omit{the formulas describe a superset of the set of program states that can actually arise.}
These methods are
employed for semi-automatic program verification (i.e., using
programmer-supplied loop invariants) on programs such as mark-and-sweep garbage
collection and destructive reversal of a singly linked list.  (The
mark-and-sweep example has been previously reported as being beyond the
capabilities of ESC/Java.)

\end{abstract}
\maketitle 
\section{\label{Se:Intro}Introduction}

This paper explores how to harness existing theorem provers for
first-order logic to prove reachability properties of programs that
manipulate dynamically allocated data structures.  The approach that
we use involves simulating reachability in a conservative way using
first-order formulas---i.e.,
 the formulas describe a superset of the set of program states
 that would be specified if one had an accurate way to express
 reachability.\eject

Automatically establishing safety and liveness properties of
sequential and concurrent programs that permit dynamic storage
allocation and low-level pointer manipulations is challenging.
Dynamic allocation causes the state space to be infinite; moreover, a
program is permitted to mutate a data structure by destructively
updating pointer-valued fields of nodes.  These features remain even if a programming language
   has good capabilities for data abstraction.
   Abstract-datatype operations are implemented
   using loops,
procedure calls, and sequences of low-level pointer manipulations;
consequently, it is hard to prove that a data-structure invariant is
reestablished once a sequence of operations is finished
\cite{kn:Hoare75}.  In languages such as Java, concurrency poses yet
another challenge: establishing the absence of deadlock requires
establishing the absence of any cycle of threads that are waiting for
locks held by other threads.

Reachability is crucial for reasoning about linked data structures.
For instance, to establish that a memory configuration contains no
garbage elements, we must show that every element is reachable from
some program variable.
Other cases where reachability is a useful notion include
\begin{enumerate}[$\bullet$]
  \item
    Specifying acyclicity of data-structure fragments, i.e.,
    from every element reachable from node $n$, one cannot reach $n$
  \item
    Specifying the effect of procedure calls when references are
    passed as arguments: only elements that are reachable
    from a formal parameter can be modified
  \item
    Specifying the absence of deadlocks
  \item
    Specifying safety conditions that allow establishing that a
    data-structure traversal terminates, e.g., there is
    a path from a node to a sink-node of the data structure.
\end{enumerate}
The verification of such properties presents a challenge.  Even simple
decidable fragments of first-order logic become undecidable when
reachability is added \cite{GOR99,eadtc}.  Moreover, the utility of
monadic second-order logic on trees is rather limited because (i)~many
programs allow non-tree data structures, (ii)~expressing the
postcondition of a procedure (which is essential for modular
reasoning) usually requires referring to the pre-state that holds before the
procedure executes, and thus cannot, in general, be expressed in
monadic second-order logic on trees---even for procedures that
manipulate only singly-linked lists, such as the in-situ list-reversal program shown in
\figref{reversecode},
and (iii)~the complexity is prohibitive.

While our work was actually motivated by our experience
using abstract interpretation -- and, in particular, the
TVLA system \cite{SAS:LS00,TOPLAS:SRW02,CAV:RSW04} --
to establish properties of programs that manipulate heap-allocated
data structures, in this paper, we consider the problem of
verifying data-structure operations, assuming that we have
user-supplied loop invariants.
This is similar to the approach taken in systems
like ESC/Java~\cite{PLDI:FLLNSS02}, and Pale~\cite{PLDI:MS01}.

The contributions of the paper can be summarized as follows:

\smallskip

\textbf{Handling \fotc\ formulas using \fo\ theorem provers.}
We want to use first-order
theorem provers and we need to discuss the transitive closure of certain binary predicates, $f$.
However, first-order theorem provers cannot handle transitive closure.
We solve this conundrum by adding a new relation symbol $\stc{f}$ for
each such $f$, together with first-order axioms that assure that
$\stc{f}$ is interpreted correctly. The theoretical details of
how this is done are presented in Section~\ref{Se:Completeness}.
The fact that we are able to
handle transitive closure effectively and reasonably automatically is quite surprising.

    As explained in \secref{Completeness}, the axioms that we add to
    control the behavior of the added predicates, $\stc{f}$, must be
    sound but not necessarily complete.  One way to think about this
    is that we are simulating a formula, $\chi$, in which transitive
    closure occurs, with a pure first-order formula $\chi'$.  If our
    axioms are not complete then we are allowing $\chi'$
     to denote  more stores than $\chi$ does.
    The study of methods that are sound but potentially incomplete is
    motivated by the fact that \emph{abstraction} \cite{POPL:CC77} can
    be an aid in the verification of many properties. In terms of logic,
    abstraction corresponds to using formulas that describe a superset
    of the set of program states that can actually arise. A definite
    answer about whether a property always holds can sometimes be
    obtained even when information has been lost because of abstraction.
 \Omit{
    This is motivated
    by the fact that abstraction can be an aid in the verification of
    many properties; that is, a definite answer can sometimes be
    obtained even when information has been lost (in a conservative
    manner).  This means that our methods are sound but potentially
    incomplete.}

    If $\chi'$ is proven valid in \fo\ then $\chi$ is also valid in
\fotc;
    however, if we fail to prove that $\chi'$ is valid, it is still
possible
    that $\chi$ is valid:  the failure would be due to the
incompleteness
    of the axioms, or the lack of time or space for the theorem prover
    to complete the proof.

As we will see in Section \ref{Se:Completeness}, it is easy to write a sound axiom, $T_1[f]$, that is ``complete'' in the very
     limited sense that every finite, acyclic model satisfying $T_1[f]$
     must interpret $\stc{f}$ as the reflexive, transitive closure of
     its interpretation of $f$.  However, in practice this is
     not worth much because, as is well-known, finiteness is not
     expressible in first-order logic.  Thus, the properties that we
     want to prove do not follow from $T_1[f]$.  We do prove that $T_1[f]$
     is complete for positive transitive-closure properties
     (Proposition \ref{Prop:positive}).  The
     real difficulty lies in proving properties involving the
     negation of $\stc{f}$, i.e., that a certain $f$-path does not exist.

 \textbf{Induction axiom scheme.}  To solve the above problem, we
 add an induction axiom scheme. Although in general, there is no
 complete, recursively-enumerable axiomatization of transitive closure
 (Proposition \ref{Prop:notre}),
 we have found, on the practical side, that on the examples we have tried,
 $T_1$ plus induction allows us to automatically prove all of our desired
 properties.
On the theoretical side,
we prove that our axiomatization is complete for word models (Theorem \ref{theorem:complete-word}).

We think of the axioms that we use as aides for the first-order
      theorem prover that we employ (\tSpass~\cite{CADE:SPASS96}) to
      prove the properties in
      question.  Rather than giving \tSpass\ many instances of
      the induction scheme, our experience is that it finds the proof faster
      if we give it several axioms that are
      simpler to use than induction.  As already mentioned,
      the hard part is to show that certain paths do not exist.

 \textbf{Coloring axiom schemes.}
      In particular, we use three axiom schemes,
      having to do with partitioning memory into a small set of
      colors.  We call instances of these schemes ``coloring axioms''.
      Our coloring axioms are simple, and are \textbf{\textit{easily proved using
        \tSpass\
      (in under ten seconds) from the induction axioms}}.  For example,
      the first
      coloring axiom scheme, $\noexit[A,f]$, says that if no $f$-edges leave
      color class, $A$, then no $f$-paths leave $A$.  It turns out that the
      $\noexit$ axiom scheme implies -- and thus is equivalent to --
      the induction scheme.  However, we have found in practice that
      explicitly adding other coloring axioms (which are consequences of
      $\noexit$) enables  \tSpass\ to prove properties that it otherwise
      fails at.

     We first assume that the programmer provides the colors by means
     of first-order
     formulas with transitive closure.  Our initial experience
     indicates that the generated coloring axioms are useful to
     \tSpass.
    In particular, it provides the ability to verify programs like
    the mark phase of a mark-and-sweep garbage collector.
    This example has been previously reported as being beyond the
    capabilities of ESC/Java.  TVLA also succeeds on this example;
    however our new approach provides verification methods that can
    in some instances be more precise than TVLA.

\textbf{Prototype implementation.}
    Perhaps most exciting, we have implemented the heuristics for
    selecting colors and their corresponding axioms in a prototype using \tSpass.
    We have used this to automatically choose useful color
    axioms and then verify a series of small heap-manipulating programs.
    We believe that the detailed examples presented here give convincing
    evidence of the promise of our methodology.  Of course much
    further study is needed.

\textbf{Strengthening Nelson's results.}
  Greg Nelson considered a set of axiom schemes for reasoning about
    reachability in function graphs, i.e., graphs in which there is at
    most one $f$-edge
    leaving any node \cite{Nelson}.  He left open the question of
    whether his axiom schemes were
    complete for function graphs.  We show that Nelson's axioms are
    provable from $T_1$
   plus our induction axioms.  We also show that Nelson's axioms are
   not complete:  in fact, they do not imply  $\noexit$.

\textbf{Outline.}
The remainder of the paper is organized as follows:
Section 2 explains our notation and the setting;
Section 3 fills in our formal framework, introduces the induction axiom scheme, and
presents the coloring axiom schemes;
Section 4 provides more detail about TC-completeness including a
description of Nelson's axioms, a proof that they are not
TC-complete for the functional case, and a proof that our axiomatization is TC-complete for words;
Section 5 presents our heuristics including the details of their
successful use on a variety of examples;
Section 6 describes the applicability of our methodology, relating
it to the reasoning done in the TVLA system;
Section 7 describes some related work; and
Section 8 describes some conclusions and future directions.

\newcommand{\qsep}{\, . \,}
\section{Preliminaries}
\label{Se:Preliminaries}

This section defines the basic notations used in this paper and the setting.

\subsection{Notation}
\subsubsubsection{Syntax} A relational \textbf{vocabulary}  $\tau= \{p_1, p_2,
\ldots, p_k\}$ is a set of relation symbols, each of fixed arity.
We use the letters $u$, $v$, and $w$ (possibly with numeric subscript) for first-order variables.
We write
first-order formulas over $\tau$ with quantifiers $\forall$ and $\exists$,
logical connectives $\land$, $\lor$, $\rightarrow$, $\leftrightarrow$, and
$\neg$, where atomic formulas include: equality, $p_i(v_1, v_2, \ldots
v_{a_i})$, and $\usetc{f}(v_1, v_2)$, where $p_i \in \tau$ is of arity $a_i$ and $f
\in \tau$ is binary. Here $\usetc{f}(v_1, v_2)$
denotes the existence of a finite path of 0 or more $f$ edges from
$v_1$ to $v_2$. A formula without $\tc$ is called a \textbf{first-order}
formula.

We use the following precedence of
logical operators: $\lnot$ has highest
precedence, followed by  $\land$ and $\lor$, followed by
$\rightarrow$ and $\leftrightarrow$, and $\forall$ and $\exists$ have
lowest precedence.

\subsubsubsection{Semantics} A \textbf{model}, $\cA$, of
vocabulary $\tau$, consists of a non-empty universe, $\abs{\cA}$,
and a relation $p^\cA$ over the universe interpreting each relation
symbol $p\in \tau$.  We write $\cA\models \phi$ to mean that the
formula $\phi$ is true in the model $\cA$.
For $\Sigma$ a set of formulas, we write $\Sigma\models \phi$
($\Sigma$ semantically implies $\phi$) to mean that all models of
$\Sigma$ satisfy $\phi$.

\subsection{Setting}

We are primarily interested in formulas that arise while proving the correctness of programs.
We assume that the programmer specifies pre and post-conditions for
procedures and loop invariants using first-order formulas with
transitive closure on binary relations.
The transformer for a loop body can be produced automatically from the program code.

 For instance, to establish the partial correctness
    with respect to a user-supplied specification
    of a program that contains a single loop, we need
    to establish three properties:
First, the loop invariant must hold at the beginning of the first iteration; i.e.,
we must show that the loop invariant follows from the precondition and the code leading to the loop.
Second, the loop
invariant provided by the user must be maintained; i.e., we must show that if the loop
invariant holds at the beginning of an iteration and the loop condition also holds,
the transformer causes the loop invariant to hold at the end of the iteration.
Finally, the postcondition must follow from the loop invariant
and the condition for exiting the loop.

In general, these formulas are of the form
\[\psi_1[\tau] \land Tr[\tau,\tau'] \rightarrow \psi_2[\tau']\]
where $\tau$ is the vocabulary of the before state, $\tau'$ is the vocabulary
of the after
state,\footnote{In some cases it is useful for the postcondition
formula to refer to the original vocabulary as well. This way the postcondition
can summarize some of the behavior of the transformer, e.g., summarize the
behavior of an entire procedure.}
and $Tr$ is the transformer, which may use both the before
and after predicates to describe the meaning of the module to be executed.  If
symbol $f$ denotes the value of a predicate before the operation, then $f'$
denotes the value of the same predicate after the operation.

An interesting special case is the proof of the maintenance formula of a loop invariant.
This  has the form:
\[LC[\tau] \land LI[\tau] \land Tr[\tau,\tau'] \rightarrow LI[\tau']\]
Here $LC$ is the condition for entering the loop and
$LI$ is the loop invariant. $LI[\tau']$ indicates that the loop
invariant remains true after the body of the loop is executed.

The challenge is that the formulas of interest contain transitive closure;
thus, the validity of these formulas cannot be directly proven using a theorem prover for first-order logic.

\section{Axiomatization of Transitive Closure}
\label{Se:Completeness}

The original formula that we want to prove, $\toprove$, contains transitive
closure, which first-order theorem provers cannot handle. To address this
problem, we replace $\toprove$ by a
new formula, $\toprove'$, where all appearances of $\usetc{f}$ have been replaced by
the new binary relation symbol, $\stc{f}$.

We show in this paper that from $\toprove'$, we can often
automatically generate an appropriate first-order axiom,
$\sigma$, with the following two properties:

\begin{enumerate}[(1)]
\item if  $\sigma \rightarrow \toprove'$ is valid in \fo, then
  $\toprove$ is valid in \fotc.
\item A theorem prover successfully proves that $\sigma \rightarrow \toprove'$ is valid in \fo.
\end{enumerate}

We now explain the theory behind this process.
A \textbf{TC model}, $\cA$, is a model such that if
  $f$ and $\stc{f}$ are in the vocabulary of $\cA$, then $(\stc{f})^\cA =
(f^\cA)^\star$; i.e., $\cA$ interprets  $\stc{f}$ as the reflexive, transitive
closure of its interpretation of $f$.

A first-order formula $\phi$ is \textbf{TC valid} iff it is true in all TC
models.  We say that an axiomatization, $\Sigma$, is \textbf{TC sound} if
every formula that follows from $\Sigma$ is TC valid.  Since
first-order reasoning is sound, $\Sigma$ is TC sound iff every $\sigma
\in \Sigma$ is TC valid.

We say that $\Sigma$ is \textbf{TC complete} if for
every TC-valid $\phi$, $\Sigma \models \phi$. If
$\Sigma$ is TC complete and TC sound, then for all first-order $\phi$,
\[\Sigma \models \phi \qLra  \phi \qquad\mbox{is TC valid }\]

Thus a TC-complete set of axioms proves exactly the first-order
formulas, $\chi'$, such that the corresponding \fotc\  formula, $\chi$, is valid.

All the axioms that we consider are TC valid.
There is no recursively enumerable TC-complete axiom system
(Proposition \ref{Prop:notre}). However, the
axiomatization that we give does allow \tSpass\ to prove all the desired
properties on the examples that we have tried.  We do prove
that our axiomatization is TC complete for word models (Theorem \ref{theorem:complete-word}).

\subsection{Some TC-Sound Axioms}

We begin with our first TC axiom scheme.  For any binary relation symbol,
$f$, let,
\[T_1[f] \qE \forall u,v \qsep \stc{f}(u,v) \slra (u=v) \lor \exists w \qsep f(u,w) \land
  \stc{f}(w,v)\]

We first observe that $T_1[f]$ is ``complete'' in a very limited way for finite,
acyclic graphs, i.e., $T_1[f]$ exactly characterizes the meaning of $\stc{f}$ for
all finite, acyclic graphs.  The reason that we say this is limited is that it does not
give us a complete set of first-order axioms: as is well known, there is
no first-order axiomatization of ``finite''.

\begin{prop}\label{Prop:limited}
Any finite and acyclic model of $T_1[f]$ is a TC model.
\end{prop}

\begin{proof}
Let $\cA \models T_1[f]$ where $\cA$ is finite and acyclic. Let $a_0,b \in
\abs{\cA}$.  Assume that there is an $f$-path from $a_0$ to $b$. Since $\cA\models
T_1[f]$, it is easy to see that $\cA\models \stc{f}(a_0,b)$.
Conversely, suppose that  $\cA\models \stc{f}(a_0,b)$.   If $a_0=b$, then there
is a path of length 0 from $a_0$ to $b$. Otherwise, by  $T_1[f]$, there exists an
$a_1 \in \abs{\cA}$ such that $\cA \models f(a_0,a_1) \land \stc{f}(a_1,b)$.
Note that $a_1 \ne a_0$ since $\cA$ is acyclic.  If $a_1 = b$ then there is an
$f$-path of length 1 from $a$ to $b$.  Otherwise there must exist an $a_2 \in
\abs{\cA}$ such that $\cA \models f(a_1,a_2) \land \stc{f}(a_2,b)$ and
so on, generating a set $\{a_1, a_2, \ldots\}$.
None of the $a_i$ can be equal to $a_j$, for $j<i$, by
acyclicity. Thus, by finiteness, some $a_i = b$.  Hence $\cA$ is a TC model.
\end{proof}

Let $\Tleft[f]$ be the $\leftarrow$ direction of $T_1[f]$:
\[\Tleft[f] \qE \forall u,v \qsep \stc{f}(u,v) \;\leftarrow\; (u=v) \lor \exists w \qsep f(u,w) \land
  \stc{f}(w,v)\]

\begin{prop}\label{Prop:positive}
Let $\stc{f}$ occur only positively in $\phi$.  If $\phi$ is TC valid,
then $\Tleft[f]\models \phi$.
\end{prop}
\begin{proof}
Suppose that $\Tleft[f]\not \models \phi $.  Let $\cA\models \Tleft[f] \land
\lnot \phi$.  Note that $\stc{f}$ occurs only negatively in $\lnot\phi$.
Furthermore, since $\cA\models \Tleft[f]$, it is easy to show by induction on
the length of the path, that if there is an $f$-path from $a$ to $b$ in $\cA$,
then $\cA\models \stc{f}(a,b)$. Define $\cA'$ to be the model formed from $\cA$
by interpreting $\stc{f}$ in $\cA'$ as $(f^\cA)^\star$.  Thus $\cA'$ is a TC
model and it only differs from $\cA$ by the fact that we have removed zero or
more pairs from $(\stc{f})^\cA$ to form  $(\stc{f})^{\cA'}$.  Because
$\cA\models \lnot \phi$ and $\stc{f}$ occurs only negatively in $\lnot \phi$,
it follows that $\cA'\models \lnot \phi$, which contradicts the assumption that
$\phi$ is TC valid.
\end{proof}

\propref{positive} shows that proving positive facts of the form $\stc{f}(u,v)$
is easy; it is the task of proving that paths do not exist that is more subtle.

\propref{limited} shows that what we are missing, at least in the
acyclic case, is that there is no first-order axiomatization of
finiteness.
Traditionally, when reasoning about the
natural numbers,  this problem is mitigated by adding induction
axioms. We next introduce an induction scheme that, together with
$T_1$, seems to be sufficient to prove any property we need concerning
TC.

\subsubsubsection{Notation}
In general, we will use $F$ to denote the set of all
binary relation symbols, $f$, such that $\usetc{f}$ occurs in a
formula we are considering.  If $\phi[f]$ is a formula in which $f$
occurs, let $\phi[F]= \bigwedge_{f\in F}\,\phi[f]$.  Thus,
for example, $T_1[F]$ is the conjunction of the axiom $T_1[f]$ for all
binary relation symbols, $f$, under consideration.

\begin{definition}
For any first-order formulas $Z(u), P(u)$, and
  binary relation symbol, $f$, let the
  \textbf{induction principle}, $\ind[Z,P,f]$, be the following first-order formula:
\begin{eqnarray*}
(\forall w\qsep  Z(w) \rightarrow P(w)) &\land& (\forall u,v \qsep P(u) \land
  f(u,v) \rightarrow P(v))\\
& \rightarrow& \forall u,w \qsep  Z(w) \land \stc{f}(w,u) \rightarrow P(u)
\end{eqnarray*}
\end{definition}

In order to explain the meaning of $\ind$ and other axioms it is
important to remember that we are trying to write axioms, $\Sigma$,
that are,
\begin{enumerate}[$\bullet$]
\item {\bf TC valid}, i.e., true in all TC models, and
\item {\bf useful}, i.e., all models of $\Sigma$ are sufficiently like
  TC models that they satisfy the TC-valid properties we want to prove.
\end{enumerate}

\noindent To make the meaning of our axioms intuitively clear, in this section
we will say, for example, that ``$y$ is $\stc{f}$-reachable from $x$''
to mean that $\stc{f}(x,y)$ holds.  Later, we will assume that the
reader has the idea and just say ``reachable'' instead of
``$\stc{f}$-reachable''.

The intuitive meaning of the induction principle is that if every zero
point satisfies $P$, and $P$ is preserved when following $f$-edges,
then every point $\stc{f}$-reachable from a zero point satisfies $P$.
Obviously this principle is TC valid, i.e., it is true for all
structures such that $\stc{f}= f^\star$.

As an easy application of the induction principle, consider the
following cousin of $T_1[f]$,
\[ T_2[f] \qE \forall u,v \qsep \stc{f}(u,v) \slra (u=v) \lor \exists w \qsep
\stc{f}(u,w) \land
  f(w,v)\]
The difference between $T_1$ and $T_2$ is that $T_1$ requires that each
path represented by $\stc{f}$ starts with an $f$ edge and $T_2$ requires
the path to end with an $f$ edge.
It is easy to see that neither of $T_1[f]$, $T_2[f]$ implies the other.
However, in the presence of the induction principle they do imply each
other.  For example, it is easy to prove $T_2[f]$ from $T_1[f]$ using
$\ind[Z,P,f]$ where $Z(v) \equiv v = u$ and $P(v) \equiv u=v  \lor
\exists w \qsep \stc{f}(u,w) \land  f(w,v)$.  Here, for each $u$ we
use $\ind[Z,P,f]$ to prove by induction that every $v$ reachable from
$u$ satisfies the right-hand side of $T_2[f]$.

Another useful axiom scheme provable from $T_1$ plus $\ind$ is
the transitivity of reachability:
\[\trans[f] \sE \forall u,v,w \qsep \stc{f}(u,w) \land
\stc{f}(w,v) \rightarrow \stc{f}(u,v)\]

\subsection{Coloring Axioms}
\label{Se:Coloring}

We next describe three TC-sound axioms schemes that are not implied by
$T_1[F] \land T_2[F]$, and are provable from the induction principle.
We will see in the sequel that these coloring axioms are very useful
in proving that paths do not exist, permitting us to verify a variety
of algorithms.  In \secref{Meth}, we will present some heuristics for
automatically choosing particular instances of the coloring axiom
schemes that enable us to prove our goal formulas.

The first coloring axiom scheme is the NoExit axiom scheme:
\[(\forall u, v  \qsep  A(u) \land \lnot A(v) \rightarrow \lnot   f(u,v))
  \qra
\forall u, v  \qsep  A(u) \land \lnot A(v) \rightarrow \lnot   \stc{f}(u,v)\]
for any first-order
formula $A(u)$, and binary relation symbol, $f$, $\AXIOM{\noexit}[A,f]$ says
that if no $f$-edge leaves color class $A$, then no point outside of $A$
is $\stc{f}$-reachable from $A$.

Observe that although it is very simple, $\AXIOM{\noexit}[A,f]$ does
not follow from $T_1[f]\land T_2[f]$.  Let $G_1 = (V,f,\stc{f},A)$
be a model consisting of two disjoint cycles: $V = \set{1,2,3,4}$, $f =
\set{\angle{1,2},\angle{2,1}, \angle{3,4}, \angle{4,3}}$, and
$A=\set{1,2}$.  Let
$\stc{f}$ have all 16 possible pairs.  Thus $G_1$ satisfies
$T_1[f]\land T_2[f]$ but violates $\AXIOM{\noexit}[A,f]$.
Even for acyclic models, $\AXIOM{\noexit}[A,f]$ does
not follow from $T_1[f]\land T_2[f]$ because there are infinite models in which
the implication does not hold (Proposition \ref{Prop:nelsonNotComplete}).

$\AXIOM{\noexit}[A,f]$ follows easily from the induction principle: if no
$f$-edges leave $A$, then induction tells us that everything $\stc{f}$-reachable from a point
in $A$ satisfies $A$.  Similarly, $\AXIOM{\noexit}[A,f]$ implies the
induction axiom, $\ind[Z,A,f]$, for any formula $Z$.

\smallskip

The second coloring axiom scheme is the GoOut axiom:
for any first-order
formulas $A(u), B(u)$, and binary relation symbol, $f$, $\AXIOM{\goout}[A,B,f]$
says that if the only $f$-edges leaving color class $A$ are to $B$, then any $\stc{f}$-path
from a point in $A$ to a point not in $A$ must pass through $B$.
\[\begin{array}{rcl}
(\forall u, v  \qsep  A(u) \land \lnot A(v) \land f(u,v)  \rightarrow B(v))
  &\rightarrow&\\
\forall u, v  \qsep  A(u) \land \lnot A(v) \land \stc{f}(u,v) & \rightarrow&
\exists w \qsep  B(w) \land \stc{f}(u,w) \land \stc{f}(w,v)
\end{array}\]
To see that $\AXIOM{\goout}[A,B,f]$  follows from the induction principle,
assume that the only $f$-edges out of $A$ enter $B$.   For any fixed $u$ in $A$, we
prove by induction that any point $v$ $\stc{f}$-reachable from $u$ is either in $A$ or
has a predecessor, $b$ in $B$, that is $\stc{f}$-reachable from $u$.

\smallskip

The third coloring axiom scheme is the $\AXIOM{\newstart}$ axiom, which is
useful in the context of dynamically changing graphs:
for any first-order
formula $A(u)$, and binary relation symbols $f$ and $g$, think of $f$ as the
previous edge relation and $g$ as the current edge relation.
$\AXIOM{\newstart}[A,f,g]$ says that if there are no new edges between $A$
nodes, then any new path, i.e., $\stc{g}$ but not $\stc{f}$, from $A$ must leave $A$ to make its change:
\[\begin{array}{rcl}
(\forall u, v  \qsep  A(u) \land A(v) \land g(u,v)  \rightarrow f(u,v))
  &\rightarrow&\\
\forall u, v  \qsep  \stc{g}(u,v) \land \lnot \stc{f}(u,v) &\rightarrow&
\exists w \qsep  \lnot A(w) \land \stc{g}(u,w) \land \stc{g}(w,v)
\end{array}\]
$\AXIOM{\newstart}[A,f,g]$ follows from the induction principle by a proof that
is similar to the proof of $\AXIOM{\goout}[A,B,f]$.

\subsubsection{Linked Lists}
The spirit behind our consideration of the coloring axioms is
similar to that found in a paper of Greg Nelson's in which he introduced a set
of reachability axioms for a functional predicate, $f$, i.e., there is at most
one $f$ edge leaving any point \cite{Nelson}.  Nelson asked whether his axiom
schemes are complete for the functional setting.  We remark that Nelson's axiom
schemes are provable from $T_1$ plus our induction principle.  However,
Nelson's axiom schemes are not complete: we constructed a functional graph
that satisfies Nelson's axioms but violates $\AXIOM{\noexit}[A,f]$
(Proposition \ref{Prop:nelsonNotComplete}).

At least one of Nelson's axiom schemes seems orthogonal to our coloring
axioms and may be useful in certain proofs.  Nelson's fifth axiom scheme states
that the points reachable from a given point are linearly ordered. The
soundness of the axiom scheme is due to the fact that $f$ is functional.
We make use of a simplified version of Nelson's ordering axiom scheme: Let
$\func[f] \equiv \forall u,v,w \qsep f(u,v) \land f(u,w) \rightarrow v = w$;
then,
\[\order[f] \sE \func[f] \rightarrow  \forall u, v, w \qsep \stc{f}(u,v) \land
\stc{f}(u,w) \sra \stc{f}(v,w) \lor \stc{f}(w,v)\]

\subsubsection{Trees}
When working with programs manipulating trees, we have a fixed set of selectors
$Sel$ and transitive closure is performed on the $down$ relation, defined as
$$\forall v_1, v_2 \qsep down(v_1,v_2) \slra \bigvee_{s \in Sel} s(v_1, v_2)$$
Trees have no sharing (i.e., the $down$ relation is injective), thus a similar axiom
to $\order[f]$ is used:
$$\forall u, v, w \qsep \stc{down}(v,u) \land \stc{down}(w,u) \sra \stc{down}(v,w) \lor \stc{down}(w,v)$$
Another important property of trees is that the subtrees below distinct children of a node are disjoint.
We use the following axioms to capture this, where $s_1 \neq s_2 \in Sel$:
$$\forall v, v_1, v_2, w \qsep \neg(s_1(v, v_1) \land s_2(v, v_2) \land \stc{down}(v_1, w) \land \stc{down}(v_2, w))$$

\section{On TC-Completeness}\label{Se:Nelson}

In this section we consider the concept of TC-Completeness in detail.
The reader anxious to see how we use our methodology is encouraged to
skim or skip this section.

We first show that there is no recursively enumerable TC-complete set
of axioms.

\begin{proposition}\label{Prop:notre}
Let $\Gamma$ be an r.e.\ set of TC-valid first-order
sentences.  Then $\Gamma$ is not TC-complete.
\end{proposition}

\begin{proof}
By the proof of Corollary 9, page 11 of \cite{eadtc},
there is a recursive procedure that, given any Turing machine $M_n$ as input,
\Omit{there is a recursive procedure that on input any Turing machine, $M_n$,}
produces a first-order formula $\phi_n$ in a vocabulary $\tau_n$
such that $\phi_n$ is TC-valid iff Turing machine, $M_n$, on input
$0$ never halts.  The vocabulary $\tau_n$ consists of the two binary
relation symbols, $E, \stc{E}$, constant symbols, $a,d$, and some
unary relation symbols.  It follows that if $\Gamma$ were
TC-complete, then it would prove all true instances of $\phi_n$ and
thus the halting problem would be solvable.
\end{proof}

\propref{notre} shows that even in the presence of only one binary
relation symbol, there is no $\RE$ TC-complete axiomatization.

In \cite{Avron}, Avron gives an elegant finite axiomatization of the natural
numbers using transitive closure, a successor relation and the binary
function symbol, ``$+$''.  Furthermore, he shows that multiplication
is definable in this language. Since the unique TC-model for Avron's
axioms is the standard natural numbers it follows that:

\begin{corollary}
Let $\Gamma$ be an arithmetic set of TC-valid first-order
sentences over a vocabulary including a binary relation symbol and a
binary function symbol (or a ternary relation symbol).
Then $\Gamma$ is not TC-complete.
\end{corollary}

In \propref{limited} we showed that any finite and acyclic model of
$T_1[f]$ is a TC model.  This can be strengthened to

\begin{proposition}
Any finite model of $T_1$ plus $\ind$ is a TC-model.
\end{proposition}

\begin{proof}
Let $\cA$ be a finite model of $T_1$ plus $\ind$.  Let $f$ be a binary
relation symbol, and let $a,b$ be
elements of the universe of $\cA$.  Since $\cA\models T_1$, if there is
an $f$ path from $a$ to $b$ then $\cA\models\stc{f}(a,b)$.

Conversely, suppose that there is no $f$ path from $a$ to $b$.  Let
$R_a$ be the set of elements of the universe of $\cA$ that are
reachable from $a$.  Let $k=\abs{R_a}$.
Since $\cA$ is finite we may use existential
quantification to name exactly all the elements of $R_a: x_1, \ldots, x_k$.
We can then define the color class: $C(y)\equiv y = x_1
\lor \cdots \lor  y = x_k$.  Then we can prove using $\ind$, or equivalently
{\bf NoExit}, that no vertex outside this color class is reachable
from $a$, i.e., $\cA\models\lnot\stc{f}(a,b)$.
Thus, as desired, $\cA$ is a TC-model.
\end{proof}

\subsection{More About TC-Completeness}

Even though there is no $\RE$ set of TC-complete axioms in general,
there are TC-complete axiomatizations for certain interesting cases.
Let $\Sigma$ be a set of formulas.  We say that $\psi$ is {\it
  TC-valid wrt} $\Sigma$ iff every TC-model of $\Sigma$ satisfies
$\psi$.  Let $\Gamma$ be TC-sound.  We say that $\Gamma$ is {\it
  TC-complete wrt} $\Sigma$ iff $\Gamma\cup\Sigma \proves \psi$ for
every $\psi$ that is TC-valid wrt $\Sigma$.
We are interested in
whether $T_1$ plus $\ind$ is TC-complete with respect to interesting
theories, $\Sigma$.

Since $\usetc{s}(a,b)$ asserts the existence of a finite $s$-path from $a$ to
$b$, we can express that a structure is finite by writing the formula: $\Phi
\equiv \func[s] \land \exists x \forall y \qsep \stc{s}(x,y)$.  Observe that
every TC-model that satisfies $\Phi$ is finite.  Thus, if we are in a setting
-- as is frequent in logic -- where we may add a new binary relation symbol,
$s$, then {\bf finiteness is TC-expressible}.

\begin{proposition}\label{Prop:Viktor}
Let $\Sigma$ be a finite set of formulas, and $\Gamma$ an r.e., TC-complete
axiomatization wrt $\Sigma$ in a language where finiteness is TC-expressible.
Then finite TC-validity for $\Sigma$ is decidable.
\end{proposition}

\begin{proof}
Let $\Phi$ be a formula as above that TC-expresses finiteness.
Let $\psi$ be any formula.  If $\psi$ is not finite TC-valid wrt $\Sigma$, then we
can find a finite TC model of $\Sigma$ where $\psi$ is false.  If $\psi$ is finite
TC-valid, then $\Gamma \cup \Sigma \proves \Phi \rightarrow \psi$, and
we can find this out by systematically generating all proofs from $\Gamma$.
\end{proof}

From \propref{Viktor} we know that we must restrict our search for
cases of TC-completeness to those where finite TC-validity is
decidable.
In particular, since the finite theory of two
functional relations is undecidable, e.g., \cite{eadtc}, we know that,

\begin{corollary}\label{functionalCase}
There are no r.e.\ TC-valid axioms for the functional case even if we restrict to
at most two binary relation symbols.
\end{corollary}

\subsection{Nelson's Axioms}

Our idea of considering transitive-closure axioms is similar in spirit to the
approach that Nelson takes \cite{Nelson}. To prove some program properties, he
introduces a set of reachability axiom schemes for a functional predicate, $f$.
By ``functional'' we mean that $f$ is a partial function:  $\func[f] \equiv
\forall u,v,w \qsep f(u,v) \land f(u,w) \rightarrow v = w$.

We remark that Nelson's axiom schemes are provable from $T_1$ plus our
induction principle.  At least two of his schemes may be useful for us to add
in our approach.  Nelson asked whether his axioms are complete for the
functional setting.  It follows from Corollary \ref{functionalCase}
that the answer is no.
We prove below that Nelson's axioms do not prove {\bf NoExit}.

Nelson's basic relation symbols are ternary. For example, he writes
``\smash{$u${\tiny \begin{tabular}{c}
$\smash f$\\[-.7ex]
$\smash{\rightarrow}$\\[-1.1ex]
$\smash{x}$
\end{tabular}}$v$}''
to mean that there is an $f$-path from $u$ to $v$ that follows no edges out of
$x$.  We encode this as, $\stc{f^x}(u,v)$, where,  for each parameter $x$ we
add a new relation symbol, $f^x$, together with the assertion: $\forall
u,v\qsep f^x(u,v) \leftrightarrow f(u,v) \land (u\ne x)$. Nelson also includes
a notation for modifying the partial function $f$.  He writes, $f^{(p)}_q$ for
the partial function that agrees with $f$ everywhere except on argument $p$
where it has value $q$.  Nelson's eighth axiom scheme asserts a basic
consistency property for this notation.  In our translation we simply assert
that  $f^{(p)}_q(u,v) \leftrightarrow (u\ne p \land f(u,v))\lor (u=p \land v =
q)$.  When we translate Nelson's eighth axiom scheme the result is
tautological, so we can safely omit it.

Using our translation, Nelson's axiom schemes are the following.

\begin{enumerate}[(N1)]\itemsep=1ex
  \item $\stc{f^x}(u,v) \slra (u=v) \lor \exists z  \qsep (f^x(u,z) \land
   \stc{f^x}(z,v))$
  \item $\stc{f^x}(u,v) \land \stc{f^x}(v,w) \rightarrow \stc{f^x}(u,w) $
  \item $\stc{f^x}(u,v) \rightarrow \stc{f}(u,v)$
  \item $\stc{f^y}(u,x) \land \stc{f^z}(u,y) \rightarrow \stc{f^z}(u,x)$
  \item $\stc{f}(u,x) \rightarrow \stc{f^y}(u,x) \lor
  \stc{f^x}(u,y)$
  \item $\stc{f^y}(u,x) \land \stc{f^z}(u,y) \rightarrow
  \stc{f^z}(x,y)$
  \item $f(x,u) \land \stc{f}(u,v) \rightarrow \stc{f^x}(u,v)$
\end{enumerate}

These axiom schemes can be proved using appropriate instances of $T_1$ and the
induction principle. Just as we showed in \propref{limited} that any finite and
acyclic model of $T_1[f]$ is a TC model, we have that,

\begin{proposition}
Any finite and functional model of Nelson's axioms is a TC-model.
\end{proposition}

\begin{proof}
Consider any finite and function model, $\cM$.  We claim that for each
$f$ and $x\in \abs{\cM}$, $(\stc{f^x})^{\cM} = ((f^x)^{\cM})^\star$.  If
there is an $f^x$ path from $u$ to $v$, then it follows from repeated
uses of (N1) that $\stc{f^x}$ holds.

If there is no $f^x$ path from $u$
to $v$ and $u$ is not on an $f$-cycle, then using (N1) we can follow
$f$-edges from $u$ to the end and prove that $\stc{f^x}$ does not
hold.

If there is no $f^x$ path from $u$
to $v$ and $u$ is on an $f$-cycle containing $x$, then
using (N1) we can follow
$f$-edges from $u$ to $x$ to prove that $\stc{f^x}(u,v)$ does not
hold.

Finally, if there is no $f$ path from $u$
to $v$ and $u$ is on an $f$-cycle, suppose for the sake of a
contradiction that $\stc{f}(u,v)$ holds.
Let $x$ be the predecessor of
$u$ on the cycle.  By N7, $\stc{f^x}(u,v)$ must hold.
However, this contradicts the previous paragraph.
\end{proof}

Axiom schemes (N5) and (N7) may be useful for us to
assert when $f$ is functional.  (N5) says that the points reachable from $u$ are
totally ordered in the sense that if $x$ and $y$ are both reachable from $u$,
then in the path from $u$ either $x$ comes first or $y$ comes first. (N7) says
that if there is an edge from $x$ to $u$ and a path from $u$ to $v$, then
there is a path from $u$ to $v$ that does not go through $x$.  This implies
the useful property that no vertex not on a cycle is reachable from a vertex on
the cycle.

We conclude this section by proving the following,

\begin{proposition}\label{Prop:nelsonNotComplete}
Nelson's axioms do not imply {\bf NoExit}.
\end{proposition}

\begin{proof}
Consider the structure $G = (V,f,\stc{f}, \stc{f^0}, \stc{f^1}, \stc{f^2},
\ldots, \stc{f^\infty}, A)$ such that $V = \N \cup \{\infty\}$, the set of
natural numbers plus a point at infinity. Let $A = \N$, i.e., the color class
$A$ is interpreted as all points except $\infty$. Define $f = \{ \angle{u,u+1}
\,|\, u \in \N\}$, i.e.,
    there is an edge from every natural number to its successor, but $\infty$ is isolated.
    However, let $\stc{f}= \{ \angle{u,v} \,|\, u \leq v \}$, i.e.,
    $G$ believes that there is a path from each natural number to
    infinity.  Similarly, for each $k\in V$,
 $\stc{f^k} = \{ \angle{u,v} | u\leq v  \land (k<u \lor v\leq k) \}$.

It is easy to check that $G$ satisfies all of Nelson's axioms.

The problem is that $G\models \lnot \noexit[A,f]$.  It follows that Nelson's
axioms do not entail $\noexit[A,f]$.  This is another proof that they are not TC complete.
\end{proof}

\subsection{TC-Completeness for Words}

In this subsection, we prove that $T_1$ plus $\ind$ is TC-complete for
words.

For any alphabet, $\Sigma$, let the vocabulary of words over $\Sigma$ be
$vocab(\Sigma) = \angle{0,max; s^2, \stc{s}^2, P_{\sigma}^1: \sigma
\in \Sigma }$ . The domain of a word model is
an ordered set of positions, and the unary relation $P_{\sigma}(x)$
expresses the presence of symbol $\sigma$ at position x. $s$ is the
successor relation over positions, and $\stc{s}$ is its transitive
closure. The constants $0$ and $max$ represent the first and last
positions in the word.  A simple axiomatization of words is $A_{\Sigma
  w}$, the conjunction of the following four statements:

\begin{enumerate}[({A}1)]\itemsep=1ex
  \item $\forall x \qsep (\lnot s(x,0) \land \lnot s(max,x) \land (x
  \neq 0 \rightarrow \exists y \qsep s(y,x)) \land (x \neq max
  \rightarrow \exists y \qsep s(x,y)))$
  \item $\forall xyz \qsep ( (s(x,y) \land s(x,z)) \lor (s(y,x) \land
  s(z,x))) \rightarrow y=z$
  \item $\forall x \qsep \stc{s}(0,x) \land \stc{s}(x,max) $
  \item $\displaystyle \forall x \qsep \bigvee_{\sigma \in \Sigma}
  (P_{\sigma}(x) \land \bigwedge_{\tau \neq \sigma} \lnot P_{\tau}(x))$
\end{enumerate}

In particular, observe that a TC-model of $A_{\Sigma w}$ is exactly a $\Sigma$ word.
Let $\Gamma=\ind \cup \{T_1\}$. We wish to prove the following:

\begin{theorem}
\label{theorem:complete-word}
$\Gamma$ is TC-complete wrt $A_{\Sigma w}$.
\end{theorem}

We first note that $\Gamma \cup \{A_{\Sigma
w}\}$ implies acyclicity: $\forall xy
\qsep  s(x,y) \rightarrow \lnot \stc{s}(y,x)$. The
proof using induction proceeds as follows: in the base case, there is
no loop at $0$. Inductively, suppose there is no loop starting at $x$,
$s(x,y)$ holds, but there is a loop at $y$, i.e., $\exists z \qsep
s(y,z) \land \stc{s}(z,y)$. Then by $T_1$ and $\ind$ we know $\exists
x' \qsep \stc{s}(z,x') \land s(x',y)$, and $\stc{s}(y,x')$.  (A2)
asserts that the in-degree of $s$ is 1, which means $x'=x$ and we have
a contradiction: $\stc{s}(y,x)$.

In
order to prove Theorem \ref{theorem:complete-word}, we need to show
that if $\phi$ is true in all TC models of $\Gamma \cup
\{A_{\Sigma w}\}$, i.e., in all words, then $\Gamma \cup \{A_{\Sigma
w}\} \proves \phi$. By the completeness of first-order logic it
suffices to show that $\Gamma \cup \{A_{\Sigma w}\} \models \phi$. We
prove the contrapositive of this in Lemma \ref{lemma:completeness}. In
order to do so, we first construct a DFA $D_\phi$ that has some
desirable properties.

\begin{lemma}
\label{lemma:mphi}

For any $\phi \in \mathcal{L}(vocab(\Sigma))$ we can build a DFA
$D_\phi = (Q_\phi, \Sigma,\delta_\phi, q_1, F_\phi) $, satisfying
the following properties:

\begin{enumerate}[\em(1)]

\item The states $q_1, q_2, \ldots q_{n}$ of $D_\phi$ are
first-order definable as formulas $q_1^1, q_2^1, \ldots
q_{n}^1$, where intuitively $q_i(x)$ will mean that $D_\phi$ is in state
$q_i$ after reading symbols at word positions $0,1, \ldots,
x$.

\item The transition function $\delta_\phi$ of $D_\phi$ is captured
by the first-order definitions of the states. That is, for all $i \le
n$, $\Gamma \cup \cA_{\Sigma w}$ semantically implies the following
two formulas for every state $q_i$:

\begin{enumerate}[\em(a)]
\item   $\displaystyle  q_i(0) \quad \leftrightarrow \quad\bigvee_{\sigma \in \Sigma,
\delta_\phi(q_1,\sigma) = q_i} P_\sigma(0)$.

\item $\displaystyle \forall u,v \qsep s(u,v) \rightarrow \Bigl(q_i(v) \quad
\leftrightarrow \quad \bigvee_{\sigma \in \Sigma, \delta_\phi(q_j,\sigma) =
q_i} (P_\sigma(v) \land q_j(u))\Bigr)$.

\end{enumerate}

\item $\displaystyle \Gamma \cup \{\cA_{\Sigma w}\} \models \phi \leftrightarrow
F(max)$, \quad  where $\displaystyle F(u) \equiv \bigvee_{q_i\in F_\phi} q_i(u)$.

\end{enumerate}

\end{lemma}

\begin{proof}

We prove properties 1, 2, and 3 while constructing
$D_\phi$ and the first-order definitions of its states by induction
on the length of $\phi$. The reward is that we get a generalized form
of the McNaughton-Papert \cite{mcnaughton} construction that works on
non-standard models.

Some subformulas of $\phi$ may have free variables, e.g., $x,y$. In the
inductive step considering such subformulas, we expand the vocabulary
of the automaton to $\Sigma' = \{x,\epsilon\} \times \{y, \epsilon\}
\times \Sigma$. We write $P_\sigma(u) \land (x = u) \land (y \ne u)$
to mean that at position $u$, symbol $\sigma$ occurs, as does $x$, but
not $y$.

\textbf{Note:} Since every structure gives a unique value to each
variable, $x$, we are only interested in strings in which $x$ occurs
at exactly one position.

For the following induction, let $\Bc $ be any model of $\Gamma
\cup \{ \cA_{\Sigma w}\}$. For the intermediate stages of
induction where some variables may occur freely, we assume that
$\Bc $ interprets these free variables.  We prove that the
formulas of properties 2 and 3 must hold in $\Bc $ at each step of
the induction.

\begin{enumerate}[\hbox to6 pt{\hfill}]
\item\noindent{\hskip-11 pt\emph{Base cases}:}\enspace$\phi$ is either $P_\sigma(x)$, $x=y$, $s(x,y)$,
  or $\stc{s}(x,y)$.\smallskip

\item\noindent{\hskip-11 pt$\phi=P_\sigma(x)$:}\enspace The automaton for $P_\sigma(x)$ and its
  state definitions are shown in Fig \ref{figure:psigma}.

\begin{figure}[h]
 \begin{minipage}[c]{0.48\columnwidth}
 \qquad\includegraphics[scale=.3]{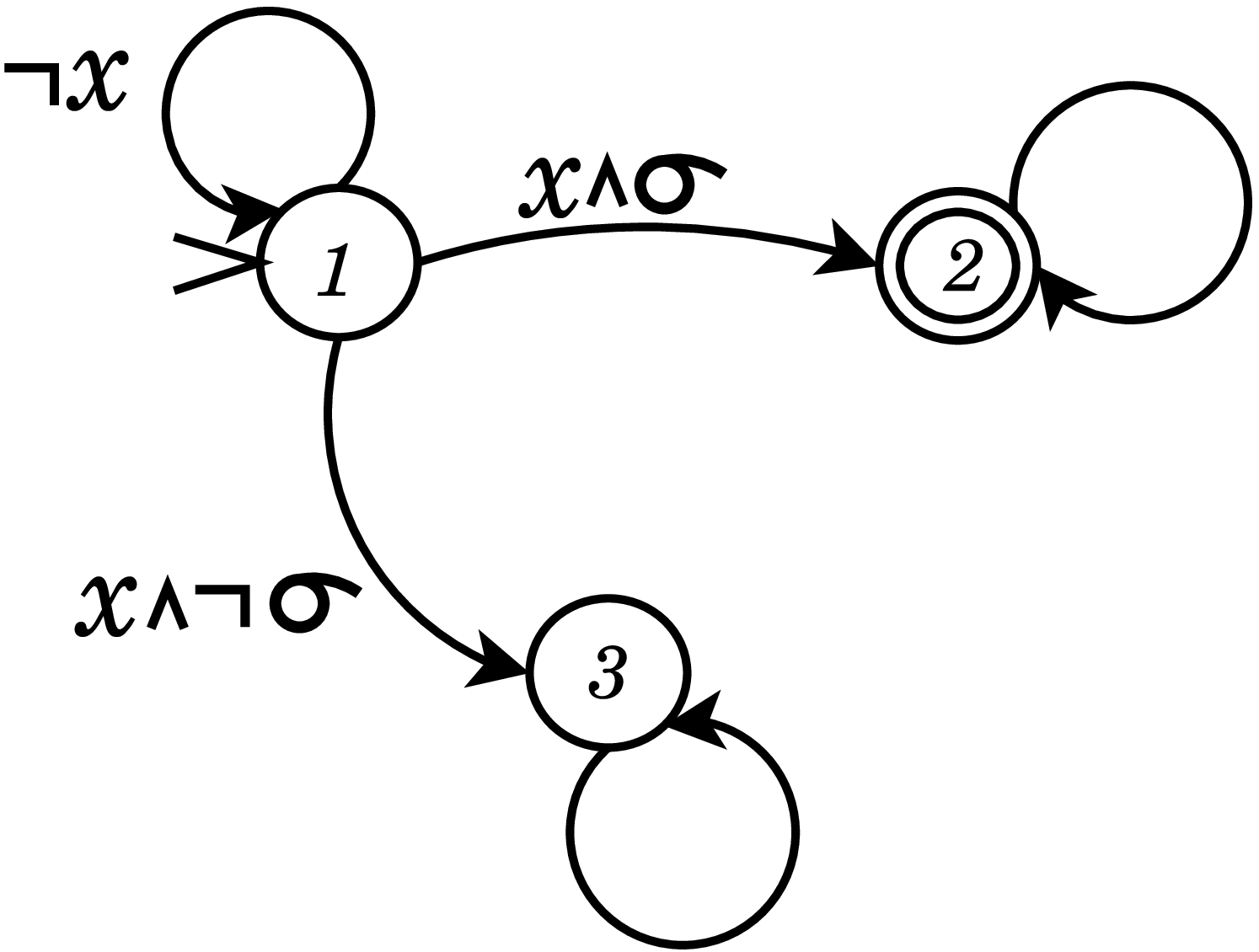}
 \caption{$D_{P_\sigma(x)}$}
 \label{figure:psigma}
 \end{minipage} \hfill
 \begin{minipage}[c]{0.48\columnwidth}
 \begin{tabular}{|c|l|}
 \hline State predicate & Definition \\
 \hline
 $q_1(v)$ & $\lnot \stc{s}(x,v)$\\
 $q_2(v)$ & $\stc{s}(x, v) \land P_\sigma(x)$ \\
 $q_3(v)$ & $\stc{s}(x, v) \land \lnot P_\sigma(x)$\\
 \hline
 \end{tabular}
 \captionof{table}{$D_{P_\sigma(x)}$}
 \label{table:psigma}
 \end{minipage}
\end{figure}

\noindent Properties 2 and 3 can be verified as follows:

For property 2b, suppose that $\Bc \models s(u,v)$. We must show that $\Bc \models q_2(v)$ iff one of two rules leading to state $q_2$
holds. These two rules correspond to the edge from $q_1$ (if $x=v$),
and the self loop on $q_2$ (if $x \ne v$). Suppose $\Bc \models q_2(v)
\land (v=x)$. Expanding the definition of $q_2$, we get
$\Bc \models \stc{s}(x,v) \land P_\sigma(x) \land (v=x)$. But this means $\Bc \models \lnot \stc{s}(x,u)$ since $\Bc
\models \Gamma \cup \{ \cA_{\Sigma w}\}$ and we have
acyclicity. Therefore, we have $\Bc \models q_1(u)$ by definition of
$q_1$, and we get the desired conclusion, $\Bc \models q_1(u) \land
P_\sigma(v)$.

The case corresponding to $x\ne v$ is also easy, and relies on the
fact that $\Bc \models \stc{s}(x,v) \land s(u,v) \land (x\ne v)
\rightarrow \stc{s}(x,u)$. In other words, if $q_2(v)$ holds and $x
\ne v$, then $q_2$ holds at $v$'s predecessor too.

This proves one direction of property 2b for state
$q_2$. The other direction for $q_2$, and the proofs for other states
proceed similarly. The proof for 2a is similar.

For property 3, we need to show that $\Bc \models P_\sigma(x)
\leftrightarrow q_2(max)$. This can be verified easily from the
definition of $q_2$.\smallskip

\item\noindent{\hskip-11 pt$\phi= (x=y)$ or $s(x,y)$:}\enspace The automata and their state
  definitions for $\phi= (x=y)$ and $\phi = s(x,y)$ are shown in Figs
  \ref{figure:xeqy} and \ref{figure:sxy}. Properties 2 and 3 can
  be verified easily for these definitions.

\begin{figure}[h]
\begin{minipage}[c]{0.48\columnwidth}
\centering
   \includegraphics[scale=.3]{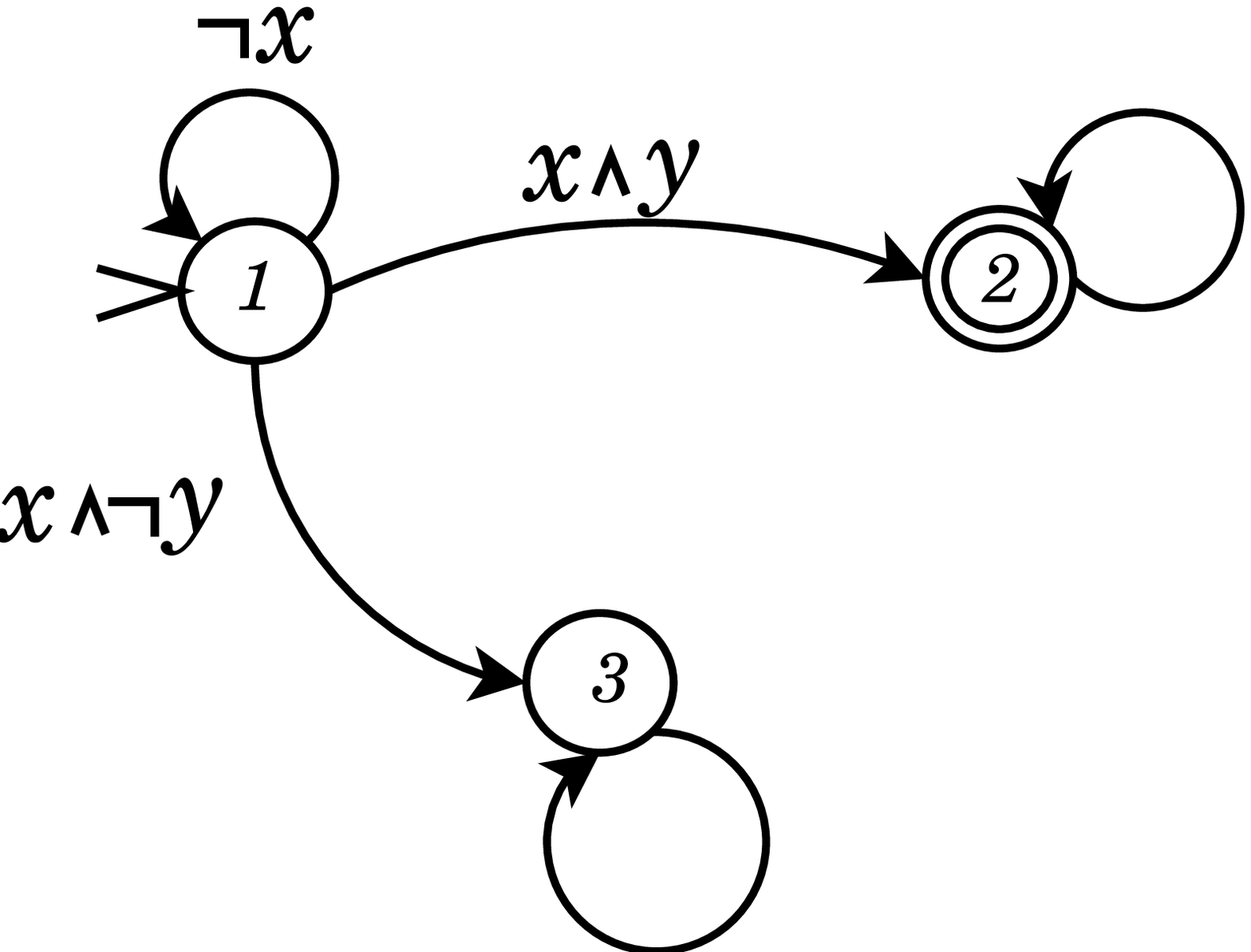}
   \captionof{figure}{$D_{x=y}$}
   \label{figure:xeqy}
\end{minipage}
\begin{minipage}[c]{0.48\columnwidth}
  \centering
  \begin{tabular}{|c|l|}
  \hline
  State predicate & Definition\\
  \hline
  $q_1(v)$ & $\lnot \stc{s}(x,v)$\\
  $q_2(v)$ & $(x=y) \land \stc{s}(x,v)$\\
  $q_3(v)$ & $(x \ne y) \land \stc{s}(x,v)$\\
  \hline
  \end{tabular}
  \captionof{table}{$D_{x=y}$}
  \label{table:xeqy}
  \end{minipage}
\end{figure}

\begin{figure}[h]
\begin{minipage}[c]{0.48\columnwidth}
\centering
  \includegraphics[scale=.3]{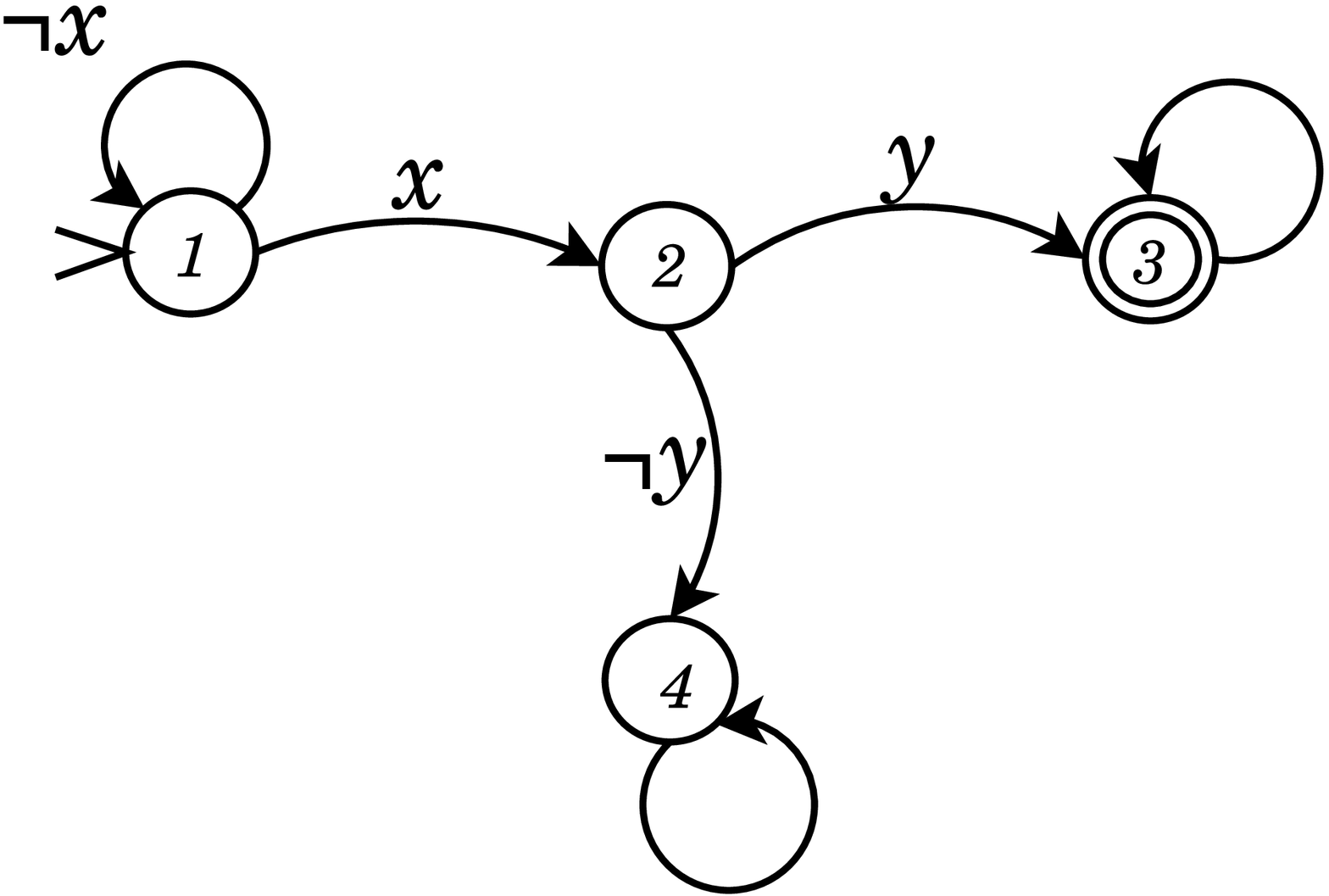}
  \captionof{figure}{$D_{s(x,y)}$}\label{figure:sxy}
\end{minipage}
  \hfill
  \begin{minipage}[c]{0.48\columnwidth}
  \centering
  \begin{tabular}{|c|l|}
  \hline
  State predicate & Definition\\
  \hline
  $q_1(v)$ & $\lnot \stc{s}(x,v)$\\
  $q_2(v)$ & $ x=v $\\
  $q_3(v)$ & $s(x,y) \land \stc{s}(y,v)$\\
  $q_4(v)$ & $\stc{s}(x,v) \land (x \ne v) \land $\\
           & $\lnot s(x,y)$\\
  \hline
  \end{tabular}
  \captionof{table}{$D_{s(x,y)}$}
  \label{table:sxy}
\end{minipage}
\end{figure}

\item\noindent{\hskip-11 pt$\phi=\stc{s}(x,y)$:}\enspace The automaton for $\phi=\stc{s}(x,y)$, and
  its state definitions are shown in Fig \ref{figure:stcxy}.

  \begin{figure}
  \begin{minipage}[c]{0.48\columnwidth}
  \qquad\includegraphics[scale=.3]{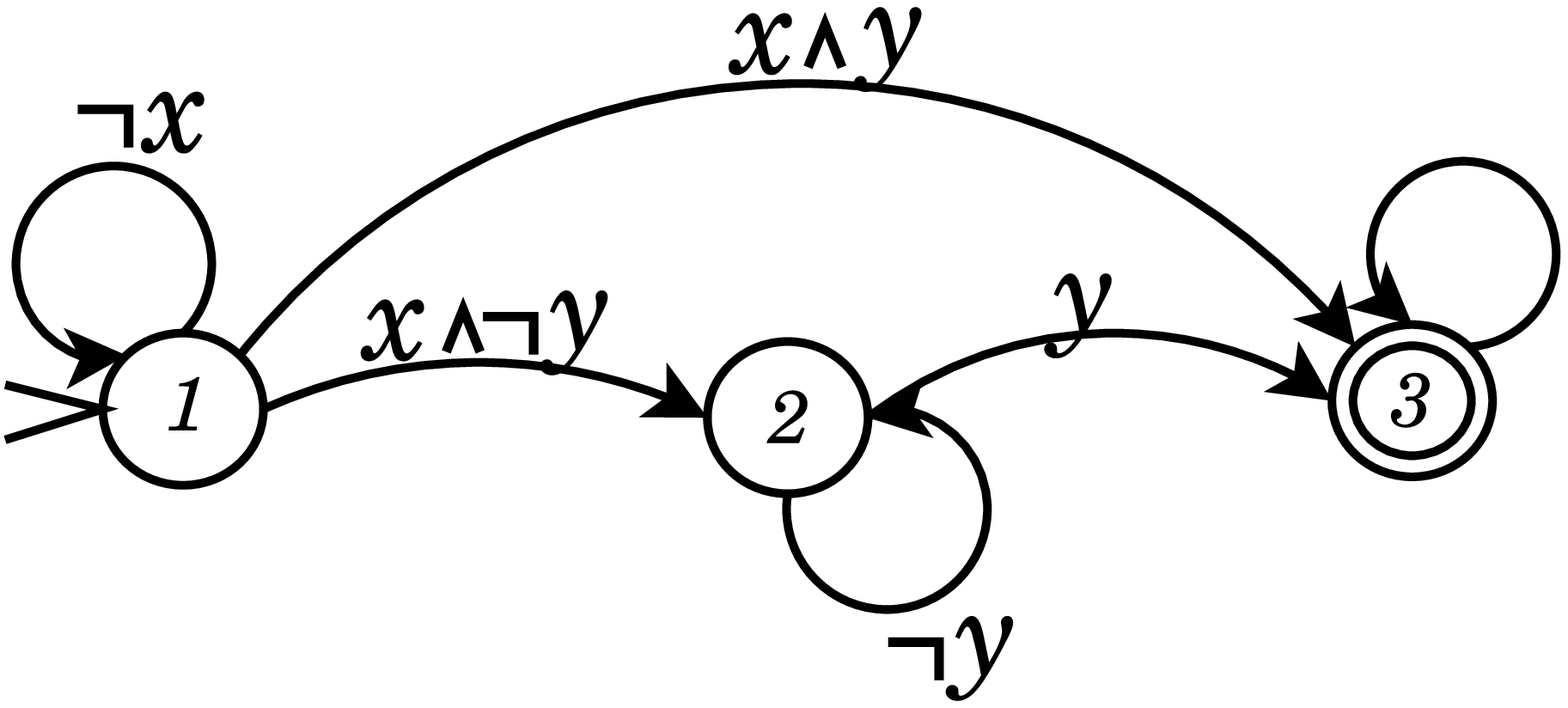}
  \captionof{figure}{$D_{\stc{s}(x,y)}$}
  \label{figure:stcxy}
  \end{minipage}
  \hfill
 \begin{minipage}[c]{0.48\columnwidth}
  \centering
  \begin{tabular}{|c|l|}
  \hline
  State predicate & Definition\\
  \hline
  $q_1(v)$ & $\lnot \stc{s}(x,v)$ \\
  $q_2(v)$ & $\stc{s}(x,v) \land $ \\
           &  $\lnot(\stc{s}(x,y) \land \stc{s}(y,v))$\\
  $q_3(v)$ & $\stc{s}(x,y) \land \stc{s}(y,v)$\\
  \hline
  \end{tabular}
  \captionof{table}{$D_{\stc{s}(x,y)}$}
  \label{table:stcxy}
  \end{minipage}
  \end{figure}

  We provide a sketch of the proof of property 2b for state
  $q_3$. Proofs for other states follow using similar
  arguments. Suppose $\Bc \models q_3(v) \land s(u,v)$.
  Expanding the definition of $q_3(v)$, we get $\Bc \models
  \stc{s}(x,y) \land  \stc{s}(y,v) \land s(u,v)$.

  There are two possibilities: $v \ne y$ and $v = y$, corresponding
  to the loop on state $q_3$, and   the incoming edges from $q_2$ or
  $q_1$. Suppose $v = y$. Now we have two further cases, $x=y$ and $x
  \ne y$. If $x=y=v$, we get   $\Bc \models \lnot \stc{s}(x,u)$, or
  $\Bc \models q_1(u) \land  s(u,x) \land (x=y=v)$, denoting the
  appropriate transition from state  $q_1$.

  On the other hand, if $\Bc \models (x \ne y)$, we
  need to show that $q_3$ was reached via $q_2$. Expanding the
  definition of $q_3(v)$ we have $\Bc \models \stc{s}(x,y) \land
  \stc{s}(y,v)$. Since $y = v$, we get $\Bc \models \stc{s}(x,u) \land
  s(u,y)$. But by definition of $q_2$, this means $\Bc \models
  q_2(u)$. Thus, we have $\Bc \models q_2(u) \land s(u,v) \land v=y$,
  the appropriate transition rule for moving from state $q_2$ to
  $q_3$.

  For this direction of property 2b, the only remaining case is $y
  \ne v$. In this case, it is easy to prove that we entered state
  $q_3$ at $y$, and looped thereafter using the appropriate transition
  for the loop.

  For the reverse direction, we need to prove that if a transition
  rule is applicable at a position then the corresponding next state
  must hold at the next position. This is easily verified using the
  state-definitions. Property 2 for other states follows by similar
  arguments. Property 3 can also be verified easily using  the
  definition of $q_3$.\smallskip

\item\noindent{\hskip-11 pt\emph{Inductive steps}:}\enspace$\phi$ is either $\phi_1 \land \phi_2$,
  or $\lnot \psi$, or $\exists x \qsep \psi(x)$.

\item\noindent{\hskip-11 pt$\phi = \phi_1 \land \phi_2$:}\enspace Inductively we have
  $D_{\phi_1}$ and $D_{\phi_2}$ with final state
  definitions $q_{f_1}$ and $q_{f_2}$ respectively. To construct
  $D_{\phi}$, we perform the product construction: let $q_i$ be
  state definitions of $D_{\phi_1}$ and $q_{i}'$ those of
  $D_{\phi_2}$. Then the state definitions of
  $D_\phi$ are $q_{\angle{ i,j}}$, and we have $q_{\angle{
  i,j}}(u) \equiv q_i(u) \land q_{j}'(u)$. The accepting states are

  \begin{displaymath} F_{\phi_1 \land \phi_2} (u) \qE \bigvee_{f_1
  \in F_1 \land f_2 \in F_2} q_{\angle{f_1, f_2} } (u).
  \end{displaymath}

  Property 1 holds because we are still in first-order. Property 2
  follows because we are just performing logical transliterations
  of the standard DFA conjunction operation. Property 3 follows from
  the fact that we already have $\Bc \models F_1(max) \leftrightarrow
  \phi_1$ and $\Bc \models F_2(max) \leftrightarrow \phi_2$, and from
  the definition of $F_{\phi_1 \land \phi_2} $.

 \item\noindent{\hskip-11 pt$\phi =\lnot \psi$:}\enspace In this case, we take the complement of
   $D_\psi$ which is easy because our automata are
   deterministic. Let the final state of $D_\psi$ be
   $F'$. $D_\phi$ has the same state definitions as $\psi$,
   but its final state definition is $F(u) \equiv \lnot F'(u)$. It is
   easy to see that properties 1, 2 and 3 hold in this case.

 \item\noindent{\hskip-11 pt$\phi = \exists x \qsep \psi(x)$:}\

 Inductively we have $D_{\psi} = (\{q_1, \ldots, q_n \}, \Sigma \times
 \{x,\epsilon\}, \delta_\psi, q_1, F_\psi)$.

 First we transform $D_\psi$ to an NFA $\Nc_\phi = (\{p_1, \ldots,
 p_n, p_1', \ldots, p_n'\}, \Sigma, \delta, p_1, F)$, where
 $F  =  \{ p_i'| q_i \in F_\psi \}$ and $\delta(p_i, \sigma) = \{
 p_j, p_k'| \delta_\psi(q_i, \sigma  \land \lnot x) = q_j,
 \delta_\psi(q_i, \sigma \land x) = q_k \}$.

 Thus $\Nc_\phi$ no longer sees $x$'s. Instead, it guesses the one
 place that $x$ might occur, and that is where the transition from $p_i$
 to $p_i'$ occurs. (See Fig. \ref{figure:psigma_lr})

 \begin{figure}[h]
 \begin{center}
 \includegraphics[scale=0.3]{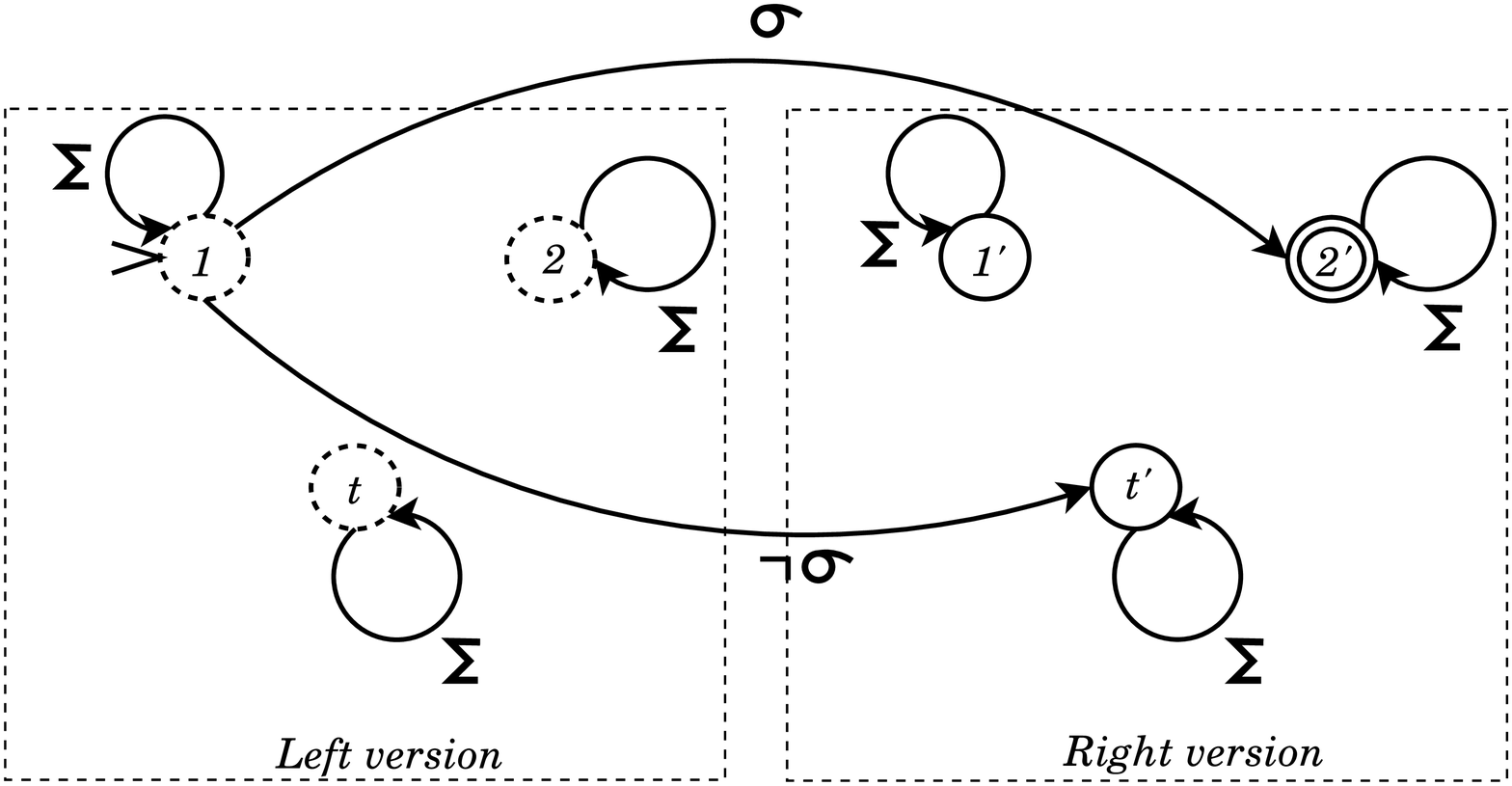}
 \caption{$\Nc_{\exists x \qsep P_\sigma(x)}$}
 \label{figure:psigma_lr}
 \end{center}
 \end{figure}

 Let $\; p_i(u)  \equiv  \exists x \qsep \lnot \stc{s}(x,u) \land q_i(u); \quad
 p_i'(u) \equiv  \exists x \qsep \stc{s}(x,u) \land q_i(u)$.

 Define $D_\phi$ to be the DFA equivalent to $\Nc_\phi$
 using the subset construction. Let $S_0 = \{p_{i_0}, p_j'| j\in J_0
 \}$, $S_1 = \{p_{i_1}, p_j'|j\in J_1 \}$ be two states of
 $D_\phi$. (Note that each reachable state of $D_\phi$ has exactly
 one element of $\{p_1, \ldots, p_n\}$.)

 Observe that in a ``run'' of $\Nc_\phi$  on $\Bc$, we can be in state
 $p_i$ at position $u$ iff $\Bc \models p_i(u)$ and we can be in state
 $p_i'$ of $u$ iff $\Bc \models p_i'(u)$. Thus, the first-order
 formula capturing state $S_0$ is
 \begin{displaymath}
 S_0(u) \qE p_{i_0} \sland \bigwedge_{j \in J_0} p_j'(u) \sland
 \bigwedge_{j \notin J_0} \lnot p_j'(u)
 \end{displaymath}

 Conditions 2 and 3 for $D_\phi$ thus follow by these conditions for
 $D_\psi$, which hold by inductive assumption.

 For example, if $\delta_\phi(S_0, \sigma) = S_1$, then
 $\delta_\psi(p_{i_0}, \sigma \land \lnot x ) = p_{i_1}$, and $j \in
 J_1$ iff $\delta_\psi(q_{i_0}, \sigma \land x) = q_j$ or
 $\delta_\psi(q_{j_0}, \sigma \land \lnot x) = q_j$ for some $j_0 \in
 J_0$.

\end{enumerate}
Thus, we have inductively constructed the $D_\phi$ and
proved that it satisfies properties 1, 2, and 3.
\end{proof}

Lemma \ref{lemma:mphi} tells us that for any model $\Bc $ of $\Gamma
\cup
\{A_{\Sigma w}\}$,  $\Bc \models \phi$ iff $\Bc \models
F_\phi(max)$. In other words, $\Bc \models \phi$ iff $\Bc $ ``believes''
that there is a path from the start state to some $q_f$ in
$F_\phi$. As a part of the next lemma, we use induction to prove that
this implies that there actually must be a path in $D_\phi$ from
the start state to some $q_f$ in $F_\phi$.

\begin{lemma}
\label{lemma:completeness}
Suppose $\Bc  \models \Gamma \cup \{A_{\Sigma w}\} \cup
\{\phi\}$. Then, there exists a word, $w_0$, such that its
corresponding word model, $\Bc _0$, satisfies $\phi$.
\end{lemma}

\begin{proof}
By Lemma \ref{lemma:mphi}, we can construct $D_\phi$, and we have
$\Bc \models F_\phi(max)$. So $\Bc$
``believes'' that there is a path to some $q_f \in F_\phi$. Suppose
there is no such path in $D_\phi$. Let $C$ denote the disjunction of
all states that are truly reachable from the start state in
$D_\phi$. This situation can be expressed as
follows: $\forall u,v \qsep C(u) \land s(u,v) \rightarrow C(v)$. But
this is exactly the premise for the axiom scheme $\noexit$, which must
hold since $\Bc \models \Gamma$. Therefore, we have
$\Bc \models \forall u, v \qsep C(v) \land \stc{s}(u,v)
\rightarrow C(v)$. This implies some accepting state $q_f$ should
be in $C$, because $\Bc \models \forall u \qsep \stc{s}(u,max) \land
F_\phi(max)$, and we get a contradiction.

Therefore, there has to be a real path from the start state to a
final state $q_f$ in $D_\phi$. This implies that the
DFA $D_\phi$ accepts some standard word, $w_0$.  Let $\Bc _0$ be
the word model corresponding to $w_0$.  Thus
$\Bc_0 \models F_\phi(max)$, and therefore by Lemma
\ref{lemma:mphi} $\Bc_0 \models \phi$ as desired.
\end{proof}


\newcommand{\Colors}{Colors}

\section{Heuristics for Using the Coloring Axioms}
\label{Se:Meth}
This section presents heuristics for using the coloring axioms.
Toward that end, it
answers the following questions:
\begin{enumerate}[$\bullet$]
\item How can the coloring axioms be used by a theorem prover to prove $\toprove$?
(\secref{Useable})
\item When should a specific instance of a coloring axiom be given to the theorem prover
while trying to prove $\toprove$? (\secref{Phased})
\item What part of the process can be automated? (\secref{Impl})
\Omit{
\item What is it about our examples that allow the method to work? When will it fail?
\item What is the price? Experimental results (number of axioms needed, time, etc.)
}
\end{enumerate}
We first present a running example (more examples are described in \secref{FullExamples}
and used in later sections to illustrate the heuristics). We then explain
how the coloring axioms are useful, describe the search space for useful
axioms, give an algorithm for exploring this space, and conclude by discussing
a prototype implementation we have developed that proves the example presented
and others.

\subsection{Reverse Specification}\label{Se:RevExample}
\label{Se:ShortExamples} The heuristics described in Sections
\ref{Se:Useable}--\ref{Se:Phased} are illustrated on problems that arise in the
verification of partial correctness of a list reversal procedure.
Other examples proven using this technique can be found in \appref{FullExamples}.

The procedure \treverse, shown in \figref{reversecode}, performs in-place reversal
of a singly linked list, destructively updating the list. The precondition
requires that the input list be acyclic and unshared
(i.e., each heap node is pointed to by at most one heap node).
For simplicity, we assume that there is no garbage. The postcondition ensures that the resulting list is
acyclic and unshared. Also, it ensures that the nodes reachable from
the formal parameter on entry to reverse are exactly the nodes reachable from
the return value of reverse at the exit. Most importantly, it ensures that each
edge in the original list is reversed in the returned list.

\begin{figure}
\begin{minipage}{3in}
\begin{alltt}
\begin{tabbing}
No\=de reverse(Node \xbefore)\+\{\\
[0] Node \ybefore = null;\\
[1] wh\=ile (\xbefore != null)\{\\
[2] \> Node t = \xbefore.next;\\
[3] \> \xbefore.next = \ybefore;\\
[4] \> \ybefore = \xbefore;\\
[5] \> \xbefore = t;\\
[6] \}\\
[7] return \ybefore;\-\\
\}
\end{tabbing}
\end{alltt}
\end{minipage}
\caption{A simple Java-like implementation of
the in-place reversal of a singly linked list.
}
\label{Fi:reversecode}
\end{figure}

The specification for \treverse\ is shown in \figref{ReverseFormula}.
We use unary predicates to represent program variables
and binary predicates to represent data-structure fields.
\figref{ReverseFormula}(a) defines some shorthands.
To specify that a unary predicate $z$ can point to a
single node at a time and that a binary predicate $f$ of a node can point to
at most one node (i.e., $f$ is a partial function), we use $unique[z]$ and $func[f]$
\Omit{, defined in \equref{unique} and \equref{func}, respectively}.
To specify that there are no cycles of $f$-fields in the graph,
we use $acyclic[f]$\Omit{, defined in \equref{acyc}}.
To specify that the graph does not contain nodes shared by $f$-fields, (i.e., nodes with $2$ or more incoming $f$-fields),
we use $unshared[f]$\Omit{, defined in \equref{unshared}}.
To specify that all nodes in the graph are reachable from $z_1$ or $z_2$ by following $f$-fields,
we use $total[z_1,z_2,f]$\Omit{, defined in \equref{total}}.
Another helpful shorthand is $r_{x,f}(v)$ which specifies that $v$ is reachable from the node pointed to by $x$ using
$f$-edges.

The precondition of the reverse procedure is shown in \figref{ReverseFormula}(b).
We use the predicates $\xentry$ and $\nentry$ to record the values of the variable $x$ and the next field at the
beginning of the procedure.
The precondition requires that the list pointed to by $\xbefore$ be acyclic and unshared.
It also requires that $unique[z]$ and $func[f]$ hold for all unary predicates $z$ that represent program variables
and all binary predicates $f$ that represent fields, respectively.
For simplicity, we assume that there is no garbage, i.e.,
all nodes are reachable from $\xbefore$.

The post-condition is shown in \figref{ReverseFormula}(c). It ensures that the
resulting list is acyclic and unshared. Also, it ensures that the nodes
reachable from the formal parameter $\xbefore$ on entry to the procedure are
exactly the nodes reachable from the return value $\ybefore$ at the exit. Most
importantly, we wish to show that each edge in the original list is reversed in
the returned list (see \equref{nReversePost}).

A loop invariant is given in \figref{ReverseFormula}(d). It describes the state
of the program at the beginning of each loop iteration. Every node is in one of
two disjoint lists pointed to by $\xbefore$ and $\ybefore$ (\equref{nxyDisjoint}).
The lists are acyclic and unshared. Every edge in the list pointed to by
$\xbefore$ is exactly an edge in the original list (\equref{xSame}). Every edge
in the list pointed to by $\ybefore$ is the reverse of an edge in the original
list (\equref{yReverse}). The only original edge going out of $\ybefore$ is to
$\xbefore$ (\equref{xy}).

The transformer is given in \figref{ReverseFormula}(e),
using the primed predicates $\nafter$, $\xafter$, and
$\yafter$ to describe the values of predicates $\nbefore$, $\xbefore$, and $\ybefore$,
respectively, at the end of the iteration.

\begin{figure}
\begin{center}
\begin{tabular}{|l@{\hspace{1pt}}l|}
\hline
(a) &
\begin{minipage}{4.8in}
\begin{eqnarray}
unique[z] & \eqdef & \forall v_1,v_2. z(v_1)  \land z(v_2) \rightarrow v_1 = v_2 \label{eq:unique}\\
func[f] & \eqdef & \forall v_1,v_2,v. f(v, v_1)  \land  f(v, v_2) \rightarrow v_1 = v_2 \label{eq:func}\\
acyclic[f] & \eqdef & \forall v_1,v_2. \lnot f(v_1, v_2)  \lor  \lnot \usetc{f}(v_2, v_1) \label{eq:acyc}\\
unshared[f] & \eqdef & \forall v_1,v_2,v. f(v_1, v) \land f(v_2, v) \rightarrow v_1 = v_2 \label{eq:unshared}\\
total[z_1, z_2,f] & \eqdef & \forall v. \exists w. (z_1(w) \lor z_2(w))  \land
\usetc{f}(w, v) \label{eq:total}\\
\FUReach{x}{f}(v) & \eqdef & \exists w \qsep x(w) \land \usetc{f}(w,v) \label{eq:UnaryReach}\\
\BUReach{x}{f}(v) & \eqdef & \exists w \qsep x(w) \land \usetc{f}(v,w) \label{eq:UnaryReachBack}
\end{eqnarray}
\end{minipage}
\\
\hline
(b) &
\begin{minipage}{4.8in}
\begin{eqnarray}
pre & \eqdef & total[\xentry, \xentry, \nentry] \land acyclic[\nentry] \land unshared[\nentry] \land
\label{eq:prec} \\
& & unique[\xentry] \land func[\nentry] \nonumber
\end{eqnarray}
\end{minipage}
\\
\hline
(c) &
\begin{minipage}{4.8in}
\begin{eqnarray}
post &\eqdef & total[\ybefore,\ybefore,\nbefore] \land acyclic[\nbefore]  \land unshared[\nbefore] \land
\label{eq:nReversePost}\\
& & \forall v_1,v_2.\nentry(v_1,v_2) \leftrightarrow \nbefore(v_2,v_1)\nonumber
\end{eqnarray}
\end{minipage}
\\
\hline
(d) &
\begin{minipage}{4.8in}
\begin{eqnarray}
LI[\xbefore, \ybefore, \nbefore] \eqdef
total[\xbefore, \ybefore, \nbefore] & \land & \forall v. (\lnot r_{\xbefore,\nbefore}(v) \lor \lnot r_{\ybefore,\nbefore}(v))
\land \label{eq:nxyDisjoint} \\
acyclic[\nbefore] & \land & unshared[\nbefore] \nonumber\\
unique[\xbefore] & \land & unique[\ybefore] \land func[\nbefore] \land \\
\forall v_1,v_2. (r_{\xbefore,\nbefore}(v_1)
& \rightarrow & (\nentry(v_1, v_2) \leftrightarrow \nbefore(v_1, v_2))) \land
\label{eq:xSame} \\
\forall v_1,v_2. (r_{\ybefore,\nbefore}(v_2) \land \lnot
\ybefore(v_1) & \rightarrow & (\nentry(v_1, v_2) \leftrightarrow \nbefore(v_2, v_1))) \land
\label{eq:yReverse} \\
\forall v_1,v_2,v. \ybefore(v_1) & \rightarrow & (\xbefore(v_2)
\leftrightarrow \nentry(v_1, v_2)) \label{eq:xy}
\end{eqnarray}
\end{minipage}
\\
\hline
(e) &
\begin{minipage}{4.8in}
\begin{eqnarray}
T \eqdef
\forall v. (\yafter(v) \leftrightarrow \xbefore(v)) & \land &
\forall v. (\xafter(v) \leftrightarrow \exists w. \xbefore(w) \land \nbefore(w, v)) \land
\nonumber\\
\forall v_1,v_2. \nafter(v_1,v_2) & \leftrightarrow &\nonumber\\
((\nbefore(v_1,v_2) \land \lnot \xbefore(v_1)) & \lor & (\xbefore(v_1)
\land \ybefore(v_2))) \label{eq:nnp}
\end{eqnarray}
\end{minipage}
\\
\hline
\end{tabular}
\end{center}
\caption{Example specification of reverse procedure:
(a)~shorthands, (b)~precondition $pre$,
(c) postcondition $post$, (d)~loop invariant $LI[\xbefore, \ybefore, \nbefore]$,
(e)~transformer $T$ (effect of the loop body).
}
\label{Fi:ReverseFormula}
\end{figure}

\subsection{Proving Formulas using the Coloring Axioms} \label{Se:Useable}
All the coloring axioms have the form $\AXIOM{A} \equiv \PREMISE{A}
\rightarrow \CONJ{A}$, where $\PREMISE{A}$ and $\CONJ{A}$ are closed formulas.
We call $\PREMISE{A}$ the axiom's premise and $\CONJ{A}$ the axiom's
conclusion. For an axiom to be useful, the theorem prover will have to prove
the premise (as a subgoal) and then use the conclusion in the proof of the goal
formula $\toprove$. For each of the coloring axioms, we now explain when the
premise can be proved, how its conclusion can help, and give an example.

\textbf{\AXIOM{\OUT}.}
The premise $\PREMISE{\oUT}[C,f]$ states that there are no $f$-edges exiting color
class $C$. When $C$ is a unary predicate appearing in the program, the premise is
sometimes a direct result of the loop invariant.
Another color that will be used heavily throughout this section is
reachability from a unary predicate, i.e., unary reachability, formally defined in \equref{UnaryReach}.
Let us examine two cases. $\PREMISE{\oUT}[r_{x,f},f]$ is immediate from the
definition of $r_{x,f}$ and the transitivity of $\stc{f}$.
$\PREMISE{\oUT}[r_{x,f},f']$ actually states that there is no $f$-path from $x$
to an edge for which $f'$ holds but $f$ does not, i.e., a change in $f'$ with respect
to $f$. Thus, we use the absence of $f$-paths to prove the absence of
$f'$-paths. In many cases, the change is an important part of the loop
invariant, and paths from and to it are part of the specification.

A sketch of the proof by refutation of $\PREMISE{\oUT}[r_{\xafter,\nbefore},\nafter]$ that arises in the
reverse example is given in \figref{rxpn}. The numbers in brackets are
the stages of the proof.
\begin{enumerate}[(1)]
\item The negation of the premise expands to: \[\exists u_1,u_2,u_3\qsep \xafter(u_1) \land \stc{n}(u_1,u_2)
\land \lnot \stc{n}(u_1,u_3) \land \nafter(u_2,u_3)\]
\item Since $u_2$ is reachable from $u_1$ and $u_3$ is not, by $T_2$, we have $\lnot n(u_2, u_3)$.
\item By the definition of $\nafter$ in the transformer, the only edge in which $\nbefore$ differs from $\nafter$
is out of $x$ (one of the clauses generated from \equref{nnp} is $\forall v_1,v_2\qsep \lnot \nafter(v_1,v_2)
\lor \nbefore(v_1,v_2) \lor \xbefore(v_1)$) . Thus, $x(u_2)$ holds.
\item By the definition of $\xafter$ it has an incoming $n$ edge from $x$. Thus, $n(u_2, u_1)$ holds.
\end{enumerate}
The list pointed to by $\xbefore$ must be acyclic, whereas we have a cycle between $u_1$ and $u_2$; i.e., we
have a contradiction. Thus, $\PREMISE{\oUT}[r_{\xafter,\nbefore},\nafter]$ must hold.

\begin{figure}
\begin{center}
\vbox{
\framebox{
\xymatrix@R10pt{
_{x'[1]}\ar[r] &
\xyNonSummaryNode{u_1}\ar[rr]^{\stc{n}[1]}\ar[dr]^{\neg \stc{n}[1]}  &
& \xyNonSummaryNode{u_2}
\ar@/_2pc/[ll]_{n[4]}\ar[dl]_{n'[1]}\ar@/^2pc/[dl]_{\neg n[2]}&
_{x [3]}\ar[l] \\
&& \xyNonSummaryNode{u_3}&
}}}
\end{center}
\caption{Proving $\PREMISE{\oUT}[r_{\xbefore,\nbefore},\nafter]$.}
\label{Fi:rxpn}
\end{figure}

$\CONJ{\oUT}[C,f]$ states there are no $f$ paths ($\stc{f}$ edges)
exiting $C$. This is useful because proving the absence of paths is the difficult part
of proving formulas with $\tc$.

\textbf{\AXIOM{\SEP}.}
The premise $\PREMISE{\sEP}[A,B,f]$ states that all $f$ edges going out of color class $A$,
go to $B$. When $A$ and $B$ are unary predicates that appear in the program, again
the premise sometimes holds as a direct result of the loop invariant. An interesting
special case is when $B$ is defined as $\exists w \qsep A(w) \land f(w,v)$. In
this case the premise is immediate. Note that in this case the conclusion is
provable also from $\NATURAL$. However, from experience, the axiom is very
useful for improving performance (2 orders of magnitude when proving the acyclic part of
\treverse's postcondition).

$\CONJ{\sEP}[A,B,f]$ states that all paths out of $A$ must pass through $B$.
Thus, under the premise $\PREMISE{\sEP}[A,B,f]$,
if we know that there is a path from $A$ to somewhere outside of $A$, we know that there is a path
to there from $B$.
In case all nodes in $B$ are reachable from all nodes in $A$, together with the
transitivity of $\stc{f}$ this means that the nodes reachable from $B$ are exactly the nodes outside of $A$
that are reachable from $A$.

For example, $\CONJ{\sEP}[\yafter,\ybefore,\nafter]$ allows us
to prove that only the original list pointed to by $\ybefore$ is
reachable from $\yafter$ (in addition to $\yafter$ itself).

\textbf{\AXIOM{\NC}.}
The premise $\PREMISE{\nC}[C,g,h]$ states that all $g$ edges between nodes in $C$ are also $h$ edges.
This can mean the iteration has not added edges or has not removed edges according to the selection
of $h$ and $g$. In some cases, the premise holds as a direct result of the definition of $C$ and the loop
invariant.

$\CONJ{\nC}[C,g,h]$ means that every $g$ path that is not an $h$ path must pass outside of $C$.
Together with $\CONJ{\oUT}[C,g]$, it proves there are no new paths within $C$.

For example, in reverse the $\AXIOM{\NC}$ scheme can be used as follows. No outgoing
edges were added to nodes reachable from $y$. There are no $\nbefore$ or $\nafter$ edges
from nodes reachable from $y$ to nodes not reachable from $y$. Thus, no paths
were added between nodes reachable from $y$. Since the list pointed to by $y$ is
acyclic before the loop body, we can prove that it is acyclic at the end of the
loop body.

We can see that $\AXIOM{\NC}$ allows the theorem prover to reason about paths within a color,
and the other axioms allow the theorem prover to reason about paths between colors. Together, given
enough colors, the theorem prover can often prove all the facts that it needs about paths and thus prove
the formula of interest.

\Omit{
 \figref{ReverseInvariantPic} shows an example state of
\treverse\ at the end of some iteration.
\begin{figure}
\begin{center}
\framebox{
\includegraphics[width=1cm,angle=-90]{rev2}
}
\end{center}
\caption{Reverse example. }
\label{Fi:ReverseInvariantPic}
\end{figure}

There is $\nbefore$-edge from $\xbefore$ to $\xafter$, but there is no
corresponding $\nafter$-edge. We say that $\xbefore$ ``covers'' this change,
because $\xbefore$ is the source of the changed edge. There is another change,
also covered by $\xbefore$: there is an $\nafter$ edge from $\xbefore$ to
$\ybefore$, but there is no corresponding $\nbefore$ edge. This change is also
covered by $\xbefore$. These are the only changes in the example.
Note that the parts of the graph that are not changed \emph{not} have \emph{no}
outgoing  $\nbefore$-edges to the changed portion of the graph. In the
subsequent sections, we show how this information help us to simulate TC.
}

\subsection{The Search Space of Possible Axioms} \label{Se:Primitive}
To answer the question of when we should use a specific instance of a coloring
axiom when attempting to prove the target formula, we first define the search
space in which we are looking for such instances. The axioms can be
instantiated with the colors defined by an arbitrary unary formula (one free
variable) and one or two binary predicates. First, we limit ourselves to binary
predicates for which $\tc$ was used in the target formula. Now, since it is
infeasible to consider all arbitrary unary formulas, we start limiting the set
of colors we consider.

The initial set of colors to consider are unary predicates that occur in the
formula we want to prove. Interestingly enough, these colors are enough to
prove that the postcondition of mark and sweep is implied by the loop
invariant, because the only axiom we need is $\AXIOM{\OUT}[marked,f]$.

An immediate extension that is very effective is
forward and backward reachability
from unary predicates, as defined in \equref{UnaryReach} and \equref{UnaryReachBack},
respectively.
Instantiating all possible axioms from the unary predicates appearing in the formula and
their unary forward reachability predicates, allows us to prove \treverse. For a list of the axioms
needed to prove \treverse, see \figref{ReverseAxioms}. Other examples are presented
in~\secref{FullExamples}.
Finally, we consider Boolean combinations of the above colors. Though not
used in the examples shown in this paper, this is needed,
for example, in the presence of sharing or when splicing two lists together.

\begin{figure}
\begin{tabular}{|@{\hspace{1ex}}l@{\hspace{1ex}}l@{\hspace{1ex}}l@{\hspace{1ex}}l|}
\hline
$\AXIOM{\OUT}[r_{\xafter, \nbefore}, \nafter]$ &
$\AXIOM{\SEP}[\xbefore, \xafter, \nbefore]$ &
$\AXIOM{\NC}[r_{\xafter, \nbefore}, \ \nbefore, \nafter]$ &
$\AXIOM{\NC}[r_{\xafter, \nbefore}, \nafter, \nbefore]$\\
$\AXIOM{\OUT}[r_{\xafter, \nafter}, \nbefore]$ &
$\AXIOM{\SEP}[\xbefore, \ybefore, \nafter]$ &
$\AXIOM{\NC}[r_{\xafter, \nafter}, \nbefore, \nafter]$ &
$\AXIOM{\NC}[r_{\xafter, \nafter}, \nafter, \nbefore]$\\
$\AXIOM{\OUT}[r_{\ybefore, \nbefore}, \nafter]$ &
&
$\AXIOM{\NC}[r_{\ybefore, \nbefore}, \ \, \nbefore, \nafter]$ &
$\AXIOM{\NC}[r_{\ybefore, \nbefore}, \nafter, \nbefore]$\\
$\AXIOM{\OUT}[r_{\ybefore, \nafter}, \nbefore]$ &
&
$\AXIOM{\NC}[r_{\ybefore, \nafter}, \nbefore, \nafter]$ &
$\AXIOM{\NC}[r_{\ybefore, \nafter}, \nafter, \nbefore]$\\
\hline
\end{tabular}
\caption{The instances of coloring axioms used in proving \treverse.}
\label{Fi:ReverseAxioms}
\end{figure}

All the colors above are based on the unary predicates that appear in the
original formula. To prove the \treverse\ example, we needed $\xafter$
as part of the initial colors. \tableref{Initial} gives a
heuristic for finding the initial colors we need in cases when they cannot be
deduced from the formula, and how it applies to \treverse.

An interesting observation is that the initial colors we need can, in many
cases, be deduced from the program code. As in the previous section, we have a
good way for deducing paths between colors and within colors in which the edges
have not changed. The program usually manipulates fields using pointers, and
can traverse an edge only in one direction. Thus, the unary predicates that
represent the program variables (including the temporary variables) are in many
cases what we need as initial colors.

\begin{table}
\begin{tabular}{l}
\begin{minipage}{5in}
\vspace{4pt}
\begin{tabular}{|l|@{\hspace{2pt}}l|}
\hline
\textbf{Group} & \textbf{Criteria}\\
\hline Roots[f] & All changes are reachable from one of
the colors using $\stc{f}$\\
 \hline
StartChange[f,g] & All edges for which $f$ and $g$ differ
 start from a node in these colors\\
 \hline
EndChange[f,g] & All edges for which $f$ and $g$ differ
 end at a node in these colors\\
 \hline
\end{tabular}
\end{minipage}
\\
(a)\\\\
\begin{minipage}{2in}
\begin{tabular}{|l|@{\hspace{10pt}}l|}
\hline
\textbf{Group} & \textbf{Colors}\\
\hline
$Roots[\nbefore]$ & $\xbefore(v)$, $\ybefore(v)$\\
\hline
$Roots[\nafter]$ & $\xafter(v)$, $\yafter(v)$\\
\hline
$StartChange[\nbefore,\nafter]$ & $\xbefore(v)$\\
\hline
$EndChange[\nbefore,\nafter]$ & $\ybefore(v)$, $\xafter(v)$\\
\hline
\end{tabular}
\end{minipage}
\\\smallskip
(b)\\
\end{tabular}
\caption{(a) Heuristic for choosing initial colors.
(b) Results of applying the heuristic on \treverse.}
\label{Ta:Initial}
\end{table}

\subsection{Exploring the Search Space}
\label{Se:Phased}

When trying to automate the process of choosing colors, the problem is that the
set of possible colors to choose from is doubly-exponential in the number of
initial colors; giving all the axioms directly to the theorem prover is
infeasible. In this section, we define a heuristic algorithm for exploring a
limited number of axioms in a directed way. Pseudocode for this algorithm is
shown in \figref{PhaseAlg}. The operator $\vdash$ is implemented as a call to a
theorem prover.

\begin{figure}
\small
\framebox{
\begin{minipage}{4in}
\begin{alltt}
\begin{tabbing}
exp\=lore(\(Init\), \(\toprove\)) \+ \{\\
Let\=\ \(\toprove = \premises \rightarrow \conclusion\)\\
\(\axioms := \{ \trans[f], \order[f]\, |\, f \in F \} \) \\
\(\axioms := \axioms \cup \{ \NATURAL[f], T_2[f]\, |\, f \in F \} \) \\
\(C := \{ r\sb{c,f}(v)\, |\, c \in Init, f \in F \}\)\\
\(C := C \cup Init \)\\
\(i := 1\)\\
for\=ever \{\+\\
\(C' := BC(i, C)\)\\
// Phase 1\\
fo\=reach \(f \in F, c\sb{s} \neq c\sb{e} \in C'\) \+\\
if\=\ \(\axioms \land \premises \vdash \PREMISE{\sEP}[c\sb{s},c\sb{e},f]\)\+\\
\(\axioms := \axioms \cup \{ \CONJ{\sEP}[c\sb{s},c\sb{e},f] \}\)\-\-\\
// Phase 2\\
foreach \(f \in F, c \in C' \) \+\\
if \(\axioms \land \premises \vdash \PREMISE{\oUT}[c,f]\)\+\\
\(\axioms := \axioms \cup \{ \CONJ{\oUT}[c,f] \}\) \-\-\\
// Phase 3\\
foreach \(\CONJ{\oUT}[c,f] \in \axioms, g \neq f \in F\)\+\\
if\=\ \(\axioms \land \premises \vdash \PREMISE{\expandafter\nC}[c,f,g]\)\+\\
\(\axioms := \axioms \cup \{ \CONJ{\nC}[c,f,g] \}\)\-\-\\
if \=\(\axioms \land \premises \vdash \conclusion\)\+ \\
return SUCCESS \-\\
\(i := i + 1\)\-\\
\}\-\\
\}
\end{tabbing}
\end{alltt}
\end{minipage}
}
 \caption{An iterative algorithm for instantiating the axiom schemes. Each
iteration consists of three phases that augment the axiom set
$\axioms$}
\label{Fi:PhaseAlg}
\end{figure}

Because the coloring axioms have the form $\AXIOM{A} \equiv \PREMISE{A}
\rightarrow \CONJ{A}$, the theorem prover must prove $\PREMISE{A}$ or the axiom
is of no use. Therefore, the pseudocode works iteratively, trying to prove
$\PREMISE{A}$ from the current $\premises \land \axioms$, and if successful it
adds $\CONJ{A}$ to $\axioms$.

The algorithm tries colors in increasing levels of complexity. $BC(i, C)$ gives
all the Boolean combinations of the predicates in $C$ up to size $i$. After
each iteration we try to prove the goal formula. Sometimes we need the
conclusion of one axiom to prove the premise of another. The $\AXIOM{\OUT}$
axioms are particularly useful for proving $\PREMISE{\nC}$. Therefore, we need
a way to order instantiations so that axioms useful for proving the premises of
other axioms are acquired first. The ordering we chose is based on phases:
First, try to instantiate axioms from the axiom scheme $\AXIOM{\SEP}$. Second,
try to instantiate axioms from the axiom scheme
$\AXIOM{\OUT}$. Finally, try to instantiate axioms from the axiom scheme
$\AXIOM{\NC}$. For $\AXIOM{\NC}[c,f,g]$ to be useful, we need to be able to
show that there are either no incoming $f$-paths or no outgoing $f$-paths from
$c$. Thus, we only try to instantiate such an axiom when either
$\PREMISE{\oUT}[c,f]$ or $\PREMISE{\oUT}[\lnot c,f]$ has been proven.

\subsection{Implementation} \label{Se:Impl}

 The algorithm presented here was implemented using a \texttt{Perl} script
and the \tSpass\ theorem prover~\cite{CADE:SPASS96} and used successfully to
verify the example programs of \secref{ShortExamples} and
\secref{FullExamples}.

The method described above can be optimized. For instance, if $\CONJ{A}$ has
already been added to the axioms, we do not try to prove $\PREMISE{A}$ again.
These details are important in practice, but have been omitted for brevity.

When trying to prove the different premises, \tSpass\ may fail to terminate if
the formula that it is trying to prove is invalid. Thus, we limit the time that
\tSpass\ can spend proving each formula. It is possible that we will fail to
acquire useful axioms this way.

\Omit{ An interesting improvement may be to integrate more closely with
\tSpass\ and try to prove the original formula in parallel with the exploration
while asserting the conclusions we can prove. }

\subsection{Further Examples}\label{Se:FullExamples}

This section shows the code (\figref{code}) and the complete specification of
two additional examples: appending two linked lists, and
the mark phase of a simple mark and sweep garbage collector.

\begin{figure}
\begin{center}
\begin{tabular}{l}
\begin{tabular}{|l|}
\hline
\begin{minipage}{4in}
\vspace{4pt}
\begin{alltt}
\begin{tabbing}
\hspace{10pt}
No\=de append(Node x, Node y)\+\ \{\\
[0] Node last = x;\\
[1] if\= (last == null)\\
[2] \> return y;\\
[3] wh\=ile (last.next != null) \{\\
[4] \> last = last.next; \\
[5] \}\\
[6] last.next = y;\\
[7] return x;\-\\
\hspace{10pt}
\}
\end{tabbing}
\end{alltt}
\end{minipage}\\
\hline
\end{tabular}
\\
(a) \\
\begin{tabular}{|l|}
\hline
\begin{minipage}{4in}
\vspace{4pt}
\begin{alltt}
\begin{tabbing}
vo\=id mark(NodeSet root, NodeSet marked)\+\ \{\\
[0] Node x;\\
[1] if\=(!root.isEmpty())\{\\
[2] \> NodeSet pending = new NodeSet();\\
[3] \> pending.addAll(root);\\
[4] \> marked.clear();\\
[5] \> wh\=ile (!pending.isEmpty())\ \{\\
[6] \>\> x = pending.selectAndRemove();\\
[7] \>\> marked.add(x);\\
[8] \>\> if\= (\=x.car != null \&\&\\
[9] \>\>\>\> !marked.contains(x.car))\\
[10] \>\>\> pending.add(x.car); \\
[11] \>\> if\= (\=x.cdr != null \&\& \\
[12] \>\>\>\> !marked.contains(x.cdr))\\
[13] \>\>\> pending.add(x.cdr); \\
\> \} \\
    \}\-\\
  \}
\end{tabbing}
\end{alltt}
\end{minipage}\\
\hline
\end{tabular}
\\
(b)\\
\end{tabular}
\end{center}
\caption{A simple Java-like implementation of
(a)~the concatenation procedure for two singly-linked lists;
(b)~the mark phase of a mark-and-sweep garbage collector.
}
\label{Fi:code}
\end{figure}

\subsubsection{Specification of append}

The specification of \tappend\ (see~\figref{code}(a)) is given in \figref{AppendFormula}.
The specification includes procedure's pre-condition, a transformer of the procedure's body effect, and
the procedure's post-condition. The pre-condition (\figref{AppendFormula}(a)) states that the lists
pointed to by $\xbefore$ and $\ybefore$ are acyclic, unshared and
disjoint. It also states there is no garbage.
The post condition (\figref{AppendFormula}(b)) states that after the procedure's execution, the list pointed to by $\xafter$ is
exactly the union of the lists pointed to by $\xbefore$ and $\ybefore$. Also, the list is still acyclic and unshared.
The transformer is given in \figref{AppendFormula}(c). The result of the loop in the procedure's body is summarized as a formula
defining the $last$ variable. The only change to $\nbefore$ is the addition of an edge between $last$ and $\ybefore$.

The coloring axioms needed to prove \tappend\ are given in
\figref{AppendAxioms}.

\begin{figure}
\begin{center}
\begin{tabular}{|cc|}
\hline
(a) &
\begin{minipage}{4.7in}
\begin{eqnarray}
pre & \eqdef & acyclic[\nbefore] \land unshared[\nbefore] \land\nonumber\\
& & unique[\xbefore] \land unique[\ybefore] \land func[\nbefore] \land\nonumber\\
& & (\forall v. \lnot r_{\xbefore,\nbefore}(v) \lor \lnot r_{\ybefore,\nbefore}(v)) \land
\forall v. r_{\xbefore,\nbefore}(v) \lor r_{\ybefore,\nbefore}(v)\label{eq:appendPRE}\\
\nonumber
\end{eqnarray}
\end{minipage}
\\
\hline
(b) &
\begin{minipage}{4.7in}
\vspace{1pt}
\begin{eqnarray}
post & \eqdef & acyclic[\nafter]  \land unshared[\nafter] \land\nonumber\\
& & unique[\xafter] \land unique[last] \land func[\nafter]\land\nonumber\\
& & (\forall v \qsep r_{\xafter, \nafter}(v) \slra (r_{\xbefore,\nbefore}(v) \lor r_{\ybefore,\nbefore}(v))) \land \nonumber\\
& & \forall v_1,v_2 \qsep \nafter(v_1,v_2) \leftrightarrow
\nbefore(v_1,v_2) \lor (last(v_1) \land \ybefore(v_2))\\
\nonumber
\end{eqnarray}
\end{minipage}
\\
\hline
(c) &
\begin{minipage}{4.7in}
\vspace{4pt}
$T$ is the conjunction of the following formulas:
\begin{eqnarray}
\forall v. \xafter(v) & \leftrightarrow & \xbefore(v)\\
\forall v. last(v) & \leftrightarrow & r_{\xbefore,\nbefore}(v) \land
\forall u. \lnot \nbefore(v,u) \label{eq:last}\\
& \exists v. & last(v) \label{eq:lastexists}\\
\forall v_1,v_2. \nafter(v_1,v_2) & \leftrightarrow &
\nbefore(v_1,v_2) \lor (last(v_1)
\land \ybefore(v_2)) \label{eq:appendnnp}
\end{eqnarray}
\end{minipage}
\\
\hline
\end{tabular}
\end{center}
\caption{Example specification of append procedure:
(a)~precondition $pre$,
(b) postcondition $post$, (c)~transformer $T$ (effect of the procedure body).
}
\label{Fi:AppendFormula}
\end{figure}

\begin{figure}
\begin{tabular}{|ll|}
\hline
$\AXIOM{\OUT}[r_{\ybefore,\nbefore},\nafter]$ &
$\AXIOM{\SEP}[last,y,\nafter]$ \\
$\AXIOM{\NC}[r_{\xbefore,\nbefore},\nbefore,\nafter]$ &
$\AXIOM{\NC}[r_{\xbefore,\nbefore},\nafter,\nbefore]$\\
$\AXIOM{\NC}[r_{\ybefore,\nbefore},\nbefore,\nafter]$ &
$\AXIOM{\NC}[r_{\ybefore,\nbefore},\nafter,\nbefore]$\\
\hline
\end{tabular}
\caption{The instances of coloring axioms used in
proving \tappend.}
\label{Fi:AppendAxioms}
\end{figure}

\subsubsection{Specification of the mark phase}
Another example proven is the mark phase of a mark-and-sweep sequential garbage
collector, shown in \figref{code}(b). The example goes beyond the reverse
example in that it manipulates a general graph and not just a linked list.
Furthermore, as far as we know, ESC/Java~\cite{PLDI:FLLNSS02} was not able
prove its correctness because it could not show that unreachable elements were not marked.
Note that the axiom needed to prove this property is $\OUT$, which we have shown to be beyond the power
of Nelson's axiomatization.

\begin{figure}
\begin{center}
\begin{tabular}{|cc|}
\hline
(a) &
\begin{minipage}{4.6in}
\begin{eqnarray}
( (\forall v\qsep  root(v) & \leftrightarrow & pending(v)) \land \\
   (\forall v\qsep  & \lnot & marked(v))) \\
    & \lor &\nonumber\\
( (\forall v\qsep  root(v) & \rightarrow & marked(v) \lor pending(v)) \land \\
   (\forall v\qsep  \lnot pending(v) & \lor & \lnot marked(v)) \land \\
   (\forall v\qsep  pending(v) & \lor & marked(v) \rightarrow
   r_{root,f}(v)) \land \\
   (\forall v_1,v_2\qsep marked(v_1) & \land & \lnot marked(v_2) \land
   \lnot pending(v_2) \nonumber\\ & \rightarrow & \lnot f(v_1,v_2)))
\end{eqnarray}
\end{minipage}
\\
\hline
(b) &
\begin{minipage}{4.6in}
\begin{eqnarray}
\forall v\qsep marked(v) \leftrightarrow r_{root,f}(v)\\
\nonumber
\end{eqnarray}
\end{minipage}
\\
\hline
\end{tabular}
\end{center}
\caption{Example specification of mark procedure:
(a) The loop invariant of mark, (b) The post-condition of mark.
}
\label{Fi:MarkFormula}
\end{figure}

The loop invariant of \begin{tt}mark\end{tt} is given in
\figref{MarkFormula}(a). The first disjunct of the formula holds only in the
first iteration, when only the nodes in root are pending and nothing is marked.
The second holds from the second iteration on. Here, the nodes in root are
marked or pending (they start as pending, and the only way to stop being
pending is to become marked). No node is both marked and pending (because the
procedure checks if the node is marked before adding it to pending). All nodes
that are marked or pending are reachable from the root set (we start with only
the root nodes as pending, and after that only nodes that are neighbors of
pending nodes became pending; furthermore, only pending nodes may become
marked). There are no edges between marked nodes and nodes that are neither
marked nor pending (because when we mark a node we add all its neighbors to
pending, unless they are marked already). Our method succeeded in proving the
loop invariant in \figref{MarkFormula}(a) using only the positive axioms.

The post-condition of \begin{tt}mark\end{tt} is given in \figref{MarkFormula}(b). To
prove it, we had to use the fact that there are no edges between marked and
unmarked nodes (i.e, there are no pending nodes at the end of the loop). Thus,
we instantiate the axiom $\OUT[marked,f]$, and this is enough to prove the
post-condition.

\section{Applicability of the Coloring Axioms}\label{Se:AbsInt}
The coloring axioms are applicable to a wide variety of verification problems.
To demonstrate this, we describe the reasoning done by the TVLA system and
how it can be simulated using the coloring axioms.
TVLA is based on the theory of abstract interpretation~\cite{kn:CC79} and
specifically on canonical abstraction~\cite{TOPLAS:SRW02}. TVLA has been
successfully used to analyze a large verity of small but intricate heap manipulating
programs (see e.g., \cite{SAS:LS00,CAV:BLRS07}), including the verification of several algorithms
(see e.g., \cite{ISSTA:LRSW00,SAS:LRS06}).
Furthermore, the axioms described in this paper have been used to integrate \tSpass\ as the
reasoning engine behind the TVLA system.
The integrated system is used to perform backward analysis on heap manipulating
programs as described in~\cite{POPL:LSR07}.

In \cite{TOPLAS:SRW02}, logical structures are used
to represent the concrete stores of the program, and \fotc\ is used to specify the concrete transformers.
This provides great flexibility in what program\-ming-language constructs
the method can handle.
For the purpose of this section, we assume that the vocabulary used is fixed
and always contains equality.
Furthermore, we assume that the transformer cannot change the universe of the
concrete store. Allocation and deallocation can be easily modeled by
using a designated unary predicate that holds for the allocated heap cells.
Similarly, we assume that the universe of the concrete store is non-empty.
Abstract stores are represented as finite $3$-valued
logical structures. We shall explain the meaning of a structure
$S$ by describing the formula $\gammaHat(S)$ to which it corresponds.

The individuals of a $3$-valued logical structure are called abstract nodes.
We use an auxiliary unary predicate for each abstract node to capture the
concrete nodes that are mapped to it. For an abstract structure with universe
$\{ node_1, \ldots, node_n \}$, let $\{ a_1, \ldots a_n \}$ be the
corresponding unary predicates.

For each $k$-ary predicate $p$ in the vocabulary,
each $k$-tuple $\tuple{node_1, \ldots, node_k}$ in the abstract structure
(called an abstract tuple) can have one of the following truth values
$\{ 0, 1, \half \}$ as follows:
\begin{enumerate}[$\bullet$]
\item The truth value $1$ means that the predicate $p$ universally holds for all
of the concrete tuples mapped to this abstract tuple, i.e.,
\begin{equation}\label{eq:tuplemust}
\forall v_1, \ldots, v_k \qsep a_1(v_1) \land \ldots \land a_k(v_k)
\rightarrow p(v_1, \ldots, v_k)
\end{equation}
\item The truth value $0$ means that the predicate $p$ universally
does not hold, for all
of the concrete tuples mapped to this abstract tuple, i.e.,
\begin{equation}\label{eq:tuplemustnot}
\forall v_1, \ldots, v_k \qsep a_1(v_1) \land \ldots \land a_k(v_k)
\rightarrow \neg p(v_1, \ldots, v_k)
\end{equation}
\item The truth value $\half$ means that we have no information about this
abstract tuple, and thus the value of the predicate $p$ is not restricted.
\end{enumerate}

We use a designated set of unary predicates called \textit{abstraction
predicates} to control the distinctions among concrete nodes
that can be made in an abstract element, which also places a bound
on the size of abstract elements.
For each abstract node $node_i$, $A_i$ denotes the set of abstraction
predicates for which $node_i$ has the truth value $1$, and
$\overline{A}_i$ denotes the set of abstraction predicates
for which $node_i$ has the truth value $0$.
Every pair $node_i, node_j$ of different abstract nodes
either $A_i \cap \overline{A}_j \neq \emptyset$ or
$\overline{A}_i \cap A_j \neq \emptyset$.
In addition, we require that the abstract nodes in the structure represent
all the concrete nodes, i.e., $\forall v\qsep \bigvee_i a_i(v)$.
Thus, the abstract nodes form a bounded partition of the concrete nodes.
Finally, each node must represent at least one concrete
node, i.e., $\exists v\qsep a_i(v)$.

The vocabulary may contain additional predicates called
\textit{derived predicates}, which are explicitly defined from other predicates
using a formula in \fotc.
These derived predicates help the precision of the analysis by recording
correlations not captured by the universal information.
Some of the unary derived predicates may also be abstraction predicates,
and thus can induce finer-granularity abstract nodes.

We say that $S_1 \sqsubseteq S_2$ if there is a total mapping $m$ between the
abstract nodes of $S_1$ and the abstract nodes of $S_2$ such that
$S_2$ represents all of the concrete stores that $S_1$ represents when
considering each abstract node of $S_2$ as a union of the
abstract nodes of $S_1$ mapped to it by $m$. Formally,
$\gammaHat(S_1) \land \psi_m \rightarrow \gammaHat(S_2)$ where
\begin{equation}
\psi_m = \bigwedge_{
\small
\begin{array}{c}
node_i \in S_1\\ m(node_i)=node'_j
\end{array}
}
\forall v\qsep a_i(v) \rightarrow a'_j(v)\nonumber
\end{equation}
The order is extended to sets using the induced Hoare order
(i.e., $\XS_1 \sqsubseteq \XS_2$ if for each element $S_1 \in \XS_1$
there exists an element $S_2 \in \XS_2$ such that $S_1 \sqsubseteq S_2$).

In the original TVLA implementation \cite{SAS:LS00} the abstract transformer
is computed by a three step process:
\begin{enumerate}[$\bullet$]
\item First, a heuristic is used to perform case splits by refining
the partition induced by the abstraction predicates. This process
is called \textit{Focus}.
\item Second, the formulas comprising the concrete transformer are used to
conservatively approximate the effect of the concrete transformer on all
the represented memory states. Update formulas are either handwritten or
derived using finite differencing~\cite{finite-differencing}.
\item Third, a constraint solver called \textit{Coerce} is used
to improve the precision of the abstract element by taking advantage
of the inter-dependencies between the predicates dictated by the
defining formulas of the derived predicates and constraints of the
programming language semantics.
\end{enumerate}

Most of the logical reasoning performed by TVLA is first order in nature.
The transitive-closure reasoning is comprised of three parts:
\begin{enumerate}[(1)]
\item The update formulas for derived predicates based on transitive closure use
first-order formulas to update the transitive-closure relation, as explained in
\secref{Precise}.
\item The Coerce procedure relates the definition of the edge relation
with its transitive closure by performing \textit{Kleene evaluation} (see below).
\item Handwritten axioms are given to Coerce to allow additional transitive-closure reasoning.
They are usually written once and for all per data-structure analyzed by the system.
\end{enumerate}

To compare the transitive-closure reasoning of TVLA and the coloring axioms presented in this
paper, we concentrate on programs that manipulate singly-linked lists and trees, although the
basic argument holds for other data-structures analyzed by TVLA as well.
The handwritten axioms used by TVLA for these cases are all covered by the axioms described in
\secref{Coloring}. The issue of update formulas is covered in detail in \secref{Precise}.
A detailed description of Kleene evaluation is beyond the scope of this paper and
can be found in \cite{TOPLAS:SRW02}. Kleene evaluation of transitive closure is equivalent
to applying transitivity to infer the existence of paths, and finding a subset of the partition
that has no outgoing edges to infer the absence of paths. The latter is equivalent to
applying the $\noexit$ axiom on the formula that defines the appropriate partition.

\subsection{Precise Update}\label{Se:Precise}
Maintenance of transitive closure through updates in the underlying
relation is required for the verification of heap-manipulating programs.
In general, it is not possible to update transitive closure for arbitrary
change using first-order-logic formulas. Instead, we limit the discussion to
unit changes (i.e., the addition or removal of a single edge).
Work in descriptive dynamic complexity~\cite{JCSS:PatnaikI1997,Hesse-thesis} and
database theory~\cite{DS95} gives first-order update formulas
to unit changes in several classes of graphs, including functional graphs
and acyclic graphs.

We demonstrate the applicability of the proposed axiom schemes by
showing how they can be used to prove the precise update formula
for unit changes in several classes of graphs.

\subsubsection{Edge addition}
We refer to the edge relation before the update by $e$ and the edge
relation after the update by $e'$.
Adding an edge from $s$ to $t$ can be formulated as
$$\forall v_1, v_2 \qsep e'(v_1, v_2) \slra (e(v_1, v_2) \lor (s(v_1) \land t(v_2))).$$
The precise update formula for this change is
$$\exists v_s, v_t \qsep s(v_s) \land t(v_t) \land
\forall v_1, v_2 \qsep \stc{e'}(v_1, v_2) \slra (\stc{e}(v_1, v_2) \lor
(\stc{e}(v_1, v_s) \land \stc{e}(v_t, v_2)))$$

We have used \tSpass\ to prove the validity of this update formula using the color axioms described
in this paper. The basic colors needed are $\FUReach{t}{e}$, i.e., forward reachability from
the target of the new edge, and $\BUReach{s}{e}$, i.e., backward reachability from the source of the
new edge. The axioms instantiated in the proof are given in \tableref{PreciseAxioms}(a).

\subsubsection{Edge removal}
There is no known precise formula for updating the transitive closure of a general graph.
For general acyclic graphs, Dong and Su~\cite{DS95} give a precise update formula that is beyond the scope of
this work. 
For functional graphs, Hesse~\cite{Hesse-thesis} gives precise update formulas based on either
an auxiliary binary relation, or by using a ternary relation to describe paths in the graph that
pass through each node. Without these additions, it is not possible to give precise update formulas
in the presence of cyclicity.

When limiting the discussion to acyclic graphs in which between any two nodes there is at most one
path (such as acyclic functional graphs and trees) it is possible to give a simple precise update
formula. As before, let $s$ be the source of the edge to be removed and $t$ be the target of the
edge. The formula for removing an edge is
$$\forall v_1, v_2 \qsep e'(v_1, v_2) \slra (e(v_1, v_2) \land \neg(s(v_1) \land t(v_2))).$$
The precise update formula for this change is
$$\exists v_s, v_t \qsep s(v_s) \land t(v_t) \land
\forall v_1, v_2 \qsep \stc{e'}(v_1, v_2) \slra (\stc{e}(v_1, v_2) \land
\neg (\stc{e}(v_1, v_s) \land \stc{e}(v_t, v_2))).$$

We have used \tSpass\ to prove the validity of this update formula
for the case of acyclic functional graphs and the case of trees.
As in edge addition, $\FUReach{t}{e}$ and $\BUReach{s}{e}$ are used as the basic colors.
The axioms instantiated in the proof are given in \tableref{PreciseAxioms}(b) and
\tableref{PreciseAxioms}(c).

\begin{table}
\begin{tabular}{ccc}
\begin{minipage}{2.2in}
\begin{tabular}{|l|}
\hline
$\NC[true,e,e']$\\
$\NC[\FUReach{t}{e} \land \neg \BUReach{s}{e}, e', e]$\\
$\NC[\neg \FUReach{t}{e} \land \BUReach{s}{e}, e', e]$\\
$\NC[\neg \FUReach{t}{e}, e', e]$\\
$\NC[\neg \BUReach{s}{e}, e', e]$\\
$\OUT[\neg \BUReach{s}{e}, e']$\\
$\OUT[\FUReach{t}{e}, e']$\\
\hline
\end{tabular}
\end{minipage}
&
\begin{minipage}{1.8in}
\begin{tabular}{|l|}
\hline
$\NC[true,e',e]$\\
$\NC[\FUReach{t}{e}, e, e']$\\
$\NC[\BUReach{s}{e}, e, e']$\\
$\NC[\neg \FUReach{t}{e}, e, e']$\\
$\NC[\neg \BUReach{s}{e}, e, e']$\\
$\OUT[\BUReach{s}{e}, e']$\\
\hline
\end{tabular}
\end{minipage}
&
\begin{minipage}{2in}
\begin{tabular}{|l|}
\hline
$\NC[true,e',e]$\\
$\NC[\FUReach{t}{e}, e, e']$\\
$\NC[\BUReach{s}{e}, e, e']$\\
$\NC[\neg \FUReach{t}{e}, e, e']$\\
$\NC[\neg \BUReach{s}{e}, e, e']$\\
$\OUT[\neg \FUReach{t}{e}, e']$\\
\hline
\end{tabular}
\end{minipage}
\\
(a) & (b) & (c)\\
\end{tabular}
\caption{\label{Ta:PreciseAxioms}Axioms instantiated for the proof of the precise update formula
of: (a)~adding an edge to a general graph,
(b)~removing an edge from an acyclic functional graph, and (c)~removing an edge from a tree.}
\end{table}

\section{Related Work}
\label{Se:RelatedWork}

\textbf{Shape Analysis.}
This work was motivated by our experience with
TVLA~\cite{SAS:LS00,TOPLAS:SRW02}, which is a generic system for abstract
interpretation~\cite{POPL:CC77}. The TVLA system is more automatic than the
methods described in this paper since it does not rely on user-supplied loop
invariants. However, the techniques presented in the present paper are
potentially more precise due to the use of full first-order reasoning. It can
be shown that the $\AXIOM{\OUT}$ scheme allows us to infer reachability at least
as precisely as evaluation rules for $3$-valued logic with Kleene semantics. In
the future, we hope to develop an efficient non-interactive theorem prover that
enjoys the benefits of both approaches.
An interesting observation is that the colors needed in our examples to prove the formula
are the same unary predicates used by TVLA to define its abstraction. This similarity may,
in the future, help us find better ways to automatically instantiate the required axioms.
In particular, inductive logic programming has recently been used to learn formulas to use
in TVLA abstractions~\cite{CAV:LRS05}, which holds out the possibility of applying similar
methods to further automate the approach of the present paper.

\textbf{Decidable Logics.}\label{Se:RelatedDecidable}
Decidable logics can be employed to define
properties of linked data structures:
Weak monadic second-order logic has been used
in~\cite{ESOP:EMS00,PLDI:MS01} to define properties of
heap-allocated data structures, and to conduct Hoare-style
verification using programmer-supplied loop invariants in the PALE system \cite{PLDI:MS01}.
A decidable logic called $L_r$ (for ``logic of
reachability expressions'') was defined in~\cite{ESOP:BRS99}.
$L_r$ is rich enough to express the shape descriptors
studied in~\cite{kn:SRW98} and the path matrices introduced
in \cite{kn:Hendren}. More recent decidable logics include Logic of Reachable Patterns
~\cite{FOSSACS:YRSMB06} and a decision procedure for linked data structures that can handle
singly linked lists~\cite{VMCAI:BR06}.

The present paper does not develop decision procedures, but instead suggests
methods that can be used in conjunction with existing theorem provers. Thus,
the techniques are incomplete and the theorem provers need not terminate.
However, our initial experience is that the extra flexibility gained by the use
of first-order logic with transitive closure is promising. For example, we can
prove the correctness of imperative destructive list-reversal specified in a
natural way and the correctness of mark and sweep garbage collectors, which are
beyond the scope of Mona and $L_r$.

Indeed, in \cite{CAV:IRRSY04}, we have tried to simulate existing data
structures using decidable logics and realized that this can be tricky because
the programmer may need to prove a specific simulation invariant for a given
program. Giving an inaccurate simulation invariant causes the simulation to be
unsound. One of the advantages of the technique described in the present paper
is that soundness is
guaranteed no matter which axioms are instantiated. Moreover, the simulation
requirements are not necessarily expressible in the decidable logic.

\textbf{Other First-Order Axiomatizations of Linked Data Structures.}
The closest approach to ours that we are aware of was taken by Nelson
as we describe in~\appref{Nelson}.  This also has some
follow-up work by Leino and Joshi~\cite{Leino}.  Our impression from
their write-up is that Leino and Joshi's work can be pushed forward by
using our coloring axioms.

A more recent work by Lahiri and Qadeer~\cite{POPL:LQ06} uses first-order axiomatization.
This work can be seen as a specialization of ours to the case of (cyclic) singly linked lists.

\textbf{Dynamic Maintenance of Transitive Closure.}
Another orthogonal but promising approach to transitive closure is to maintain
reachability relations incrementally as we make unit changes in the
data structure.  It is known that in many cases, reachability can be
maintained by first-order formulas \cite{DS95,JCSS:PatnaikI1997} and even
sometimes by quantifier-free formulas \cite{Hesse-thesis}.
Furthermore, in these cases, it is often possible to automatically
derive the first-order update formulas using finite differencing
\cite{finite-differencing}.

\section{Conclusion}
\label{Se:Conclusion}

This paper reports on our proposal of a new methodology 
for using off-the-shelf first-order theorem provers to reason about
reachability in programs.  We have explored many of the theoretical
issues as well as presenting examples that, while still preliminary,
suggest that this is indeed a viable approach. 

As mentioned earlier, proving the absence of paths is the difficult part of
proving formulas with $\tc$.  The promise of our approach is that it is able to
handle such formulas effectively and reasonably automatically, as shown by the
fact that it can successfully handle the programs described in \secref{Meth}
and the success of the TVLA system, which uses similar transitive-closure reasoning.
Of course, much further work is needed including the following:

\begin{enumerate}[$\bullet$]
  \item
    Exploring other heuristics for identifying color classes.
  \item
    Exploring variations of the algorithm given in \figref{PhaseAlg}
    for instantiating coloring axioms.
  \item
    Exploring the use of additional axiom schemes, such as two of the
    schemes from \cite{Nelson}, which are likely to be useful when
    dealing with predicates that are partial functions.  Such
    predicates arise in programs that manipulate singly-linked or
    doubly-linked lists---or, more generally, data structures that
    are acyclic in one or more ``dimensions'' \cite{kn:HHN92} (i.e.,
    in which the iterated application of a given field selector can
    never return to a previously visited node).
    \item 
    Additional work should be done on the theoretical power of $T_1 +
    \ind$ and related axiomatizations of transitive closure.  We
    conjecture, for example, that $T_1 +
    \ind$ is TC-complete for trees.
\end{enumerate}

\subsection*{Acknowledgements}
Thanks to Aharon Abadi and Roman Manevich for interesting suggestions.
Thanks to Viktor Kuncak for useful conversations including his observation and proof of
\propref{Viktor}.


  \bibliography{df,logic,more,mab,static,paper}
  \bibliographystyle{alpha}
\end{document}